\documentclass[2p]{elsarticle}

\usepackage{graphicx}
\usepackage{epsfig}
\usepackage{color}
\usepackage{psfrag}
\usepackage{subfigure}
\usepackage{caption}
\usepackage[colorlinks]{hyperref}
\usepackage{xcolor,colortbl}
\usepackage{diagbox}

\usepackage{subeqnarray}
\usepackage[ruled]{algorithm2e}
\usepackage{rotating}
\usepackage{float}

\graphicspath{{./}}
\usepackage{amsmath,amsfonts,amsfonts}
\usepackage{lineno}
\usepackage[version=3]{mhchem}
\usepackage{bm}
\usepackage{bbm}
\usepackage{amsthm}
\usepackage{dashrule}
\usepackage[titletoc,title]{appendix}

\newcommand{\wh}[1]{{\widehat{#1}}}

\journal{Combust. Flame}

\begin{document}
\newcommand{\colblue}[1]{{\color{blue} #1}}
\newcommand\norm[1]{\left\lVert#1\right\rVert}
\newcommand\chem[1]{\ce{#1}}
\def\Yvec{\boldmath{Y}}
\def\uvec{\boldmath{u}}
\def\xvec{\boldmath{x}}
\def\psivec{\boldsymbol{\psi}}
\def\omegavec{\boldsymbol{\omega}}
\def\alphavec{\boldsymbol{\alpha}}

\def\cD{{\cal{D}}}
\def\cG{{\cal{G}}}
\def\cM{{\cal{M}}}
\def\cF{{\cal{F}}}
\def\cW{{\cal{W}}}
\def\YM{{\hat{\boldsymbol\phi}}}
\providecommand{\e}[1]{\ensuremath{\times 10^{#1}}}
\newcommand{\ol}[1]{{\overline{#1}}}
\newcommand{\wt}[1]{{\widetilde{#1}}}
\def\tauvec{{\boldsymbol{\tau}}}
\newcommand{\f}[2]{{\frac{#1}{#2}}}

\begin{frontmatter}
\title{A General Probabilistic Approach for Quantitative Assessment of LES Combustion Models}
\author{Ross Johnson}
\ead{rjohnso2@stanford.edu}
\author{Hao Wu}
\ead{wuhao@stanford.edu}
\author{Matthias Ihme\corref{CORR1}}
\ead{mihme@stanford.edu}
\address{Department of Mechanical Engineering, Stanford University, Stanford, CA 94305}
\cortext[CORR1]{Corresponding author}
\begin{abstract}
The Wasserstein metric is introduced as a probabilistic method to enable quantitative evaluations of LES combustion models. The Wasserstein metric can directly be evaluated from scatter data or statistical results using probabilistic reconstruction against experimental data. The method is derived and generalized for turbulent reacting flows, and applied to validation tests involving the Sydney piloted jet flame. It is shown that the Wasserstein metric is an effective validation tool that extends to multiple scalar quantities, providing an objective and quantitative evaluation of model deficiencies and boundary conditions on the simulation accuracy. Several test cases are considered, beginning with a comparison of mixture-fraction results, and the subsequent extension to reactive scalars, including temperature and species mass fractions of \ce{CO} and \ce{CO2}. To demonstrate the versatility of the proposed method in application to multiple datasets, the Wasserstein metric is applied to a series of different simulations that were contributed to the TNF-workshop. Analysis of the results allowed to identify competing contributions to model deviations, arising from uncertainties in the boundary conditions and model deficiencies. These applications demonstrate that the Wasserstein metric constitutes an easily applicable mathematical tool that reduce multiscalar combustion data and large datasets into a scalar-valued quantitative measure.
\end{abstract}

\end{frontmatter}

\tableofcontents
\section{\label{SEC_INTRODUCTION}Introduction}
A main challenge in the development of models for turbulent reacting flows is the objective and quantitative evaluation of the agreement between experiments and simulations. This challenge arises from the complexity of the flow-field data, involving different thermo-chemical and hydrodynamic quantities that are typically provided from measurements of temperature, speciation, and velocity. This data is obtained from various diagnostics techniques, including nonintrusive methods such as laser spectroscopy and particle image velocimetry, or intrusive techniques such as exhaust-gas sampling or thermocouple measurements~\cite{HEITOR_MOREIRA_PECS1993,ECKBRETH_BOOK1996,KOHSE_HOINGHAUS_BARLOW_ALDEN_WOLFRUM_PCI2005,BARLOW_PCI2007}. Instead of directly measuring physical quantities, they are typically inferred from measured signals, introducing several correction factors and uncertainties~\cite{BARLOW_FRANK_KARPETIS_CHEN_CF2005}. Depending on the experimental technique, these measurements are generated from single-point data, line measurements, line-of-sight absorption, or multidimensional imaging at acquisition rates ranging from single-shot to high-repetition rate measurements to resolve turbulent dynamics~\cite{ALDEN_BOOD_LO_RICHTER_PCI2011}. This data is commonly processed in the form of statistical results from Favre and Reynolds averaging, conditional data, probability density functions, and scatter data. These data provide important information for model evaluation.

Significant progress has been made in simulating turbulent flows. This progress can largely be attributed to adopting high-fidelity large-eddy simulation (LES) for the prediction of unsteady turbulent flows, and establishing forums for collaborative comparisons of benchmark flames that are supported by comprehensive databases~\cite{BARLOWTNF}.

In spite of the increasing popularity of LES, the validation of combustion models follows previous steady-state RANS-approaches, comparing statistical moments (typically mean and root-mean-square) and conditional data along axial and radial locations in the flame. Qualitative comparisons of scatter data are commonly employed to examine whether a particular combustion model is able to represent certain combustion-physical processes in composition space that, for instance, are associated with extinction, equilibrium composition, or mixing conditions. In other investigations, error measures were constructed by weighting moments and scalar quantities for evaluating the sensitivity of model coefficients in subgrid models and for uncertainty quantification~\cite{KEMPF_GEURTS_OEFELEIN_CF2011,KHALIL_LACAZE_OEFELEIN_NAJM_PCI2015}. It is not uncommon that models match particular measurement quantities (such as temperature or major species products) in certain regions of the flame, while mispredicting the same quantities in other regions or showing disagreements for other flow-field data at the same location. Further, the comparison of individual scalar quantities makes it difficult to consider dependencies and identify correlations between flow-field quantities. 
Faced with this dilemma, such comparisons often only provide an inconclusive assessment of the model performance, and limit a quantitative comparison among different modeling approaches. Therefore, a need arises to develop a metric that enables a quantitative assessment of combustion models, fulfilling the following requirements:
\begin{enumerate}
 \item Provide a single metric for quantitative model evaluation;
 \item Combine single and multiple scalar quantities in the validation metric, including temperature, mixture fraction, species mass fractions, and velocity;
 \item Incorporate scatter data from single-shot, high-speed, and simultaneous measurements, and enable the utilization of statistical data;
 \item Enable the consideration of dependencies between measurement quantities; and 
 \item Ensure that metric satisfies conditions on non-negativity, identity, symmetry, and triangular inequality.
\end{enumerate}
By addressing the requirements, the objective of this work is to introduce the probabilistic Wasserstein distance~\cite{BOGACHEV_BOOK2007} as a metric for the quantitative evaluation of combustion models. Complementing currently employed comparison techniques, this metric possesses the following attractive properties. First, this metric directly utilizes the abundance of data from unsteady simulation techniques and high repetition rate measurement methods. Rather than considering lower-order statistical moments, this metric is formulated in distribution space. As such, it is thereby directly applicable to scatter data that are obtained from transient simulations and high-speed measurements without the need for data reduction to low-order statistical moment information. Second, this metric is able to synthesize multidimensional data into a scalar-valued quantity, thereby aggregating model discrepancies for individual quantities. Third, the resulting metric utilizes a normalization, thereby enabling the objective comparison of different simulation approaches. Fourth, this method is directly applicable to sample data that are generated from scatter plots, instantaneous simulation results or reconstructed from statistical results, and enables the consideration of conditional and multiscalar data. Fifth, the Wasserstein metric is equipped with essential properties of metric spaces. Finally, this metric is a companion tool to previously established methods for validation of LES and instantaneous CFD-simulations.

The remainder of this manuscript is structured as follows. Section~\ref{SEC_METHOD} formally introduces the Wasserstein metric and builds a mathematical foundation for this method. This method is demonstrated in application to the Sydney piloted jet flame with inhomogeneous inlet, and experimental configuration and simulation setup are described in Section~\ref{SEC_CONFIG}. Results are presented in Section~\ref{SEC_RESULTS}. To introduce the metric, we first consider a scalar distribution of mixture fraction and establish a quantitative comparison of experiments and simulations against presumed closure models. This is then extended to multiscalar data, involving temperature and species mass fractions. Subsequently, we examine the utility of applying the Wasserstein metric to statistical results, rather than scatter data, while Section~\ref{SEC_TNF} explores how the Wasserstein metric could add value for directly comparing multiple simulations to assess LES closure models. This work finishes in Section~\ref{SEC_CONCLUSION} with conclusions.

\section{\label{SEC_METHOD}Methodology}
In this section, the methodology of quantifying the dissimilarity between two multivariate distributions (joint PDFs) is discussed. Since multivariate distributions are commonly used to represent the thermo-chemical states in turbulent flows, this method is useful for the quantitative assessment of differences between a numerical simulation and experimental data of turbulent flames.

The dissimilarity between two points in the thermo-chemical state space can typically be quantified by its Euclidean distance. In the case of a 1D state space, e.g. temperature, the Euclidean distance is simply the absolute value of the difference. This metric serves the purpose well for the comparison among two deterministic measurements, which can indeed be fully represented by points in the state space. However, the Euclidean distance falls short when measurements are made on random data, e.g. states in a turbulent flow, in which case a single data point can no longer represent the measurement and shall be replaced by a distribution of possible outcomes. Consequently, the dissimilarity between two random variables shall be quantified by the ``distance'' between the corresponding probability distributions. 

Among many definitions of the ``distance" between distributions~\cite{gibbs2002choosing, Rubner2000}, the Wasserstein metric (also known as the Mallows metric) is of interest in this study. Many other methods of quantifying the difference between two distributions are either designed for equality test (e.g.~Kolmogorov-Smirnov test) or fails to satisfy the metric properties (e.g.~Kullback-Leibler divergence). Interested readers are referred to the survey by Gibbs and Su~\cite{gibbs2002choosing} for the detailed comparison between the Wasserstein metric and other candidates. The Wasserstein metric represents a natural extension of the Euclidean distance, which can be recovered by the Wasserstein metric as the distribution reduces to a Dirac delta function. As such, this metric provides a measure of the difference between distributions that resembles the classical idea of ``distance". The resulting value can be judged and interpreted in the similar fashion as the arithmetic difference between two deterministic quantities.

The Wasserstein metric has been used as a tool for measuring the ``distance" between distributions or histograms in the context of content-based image retrieval~\cite{Rubner2000}, hand-gesture recognition~\cite{ren2011robust}, and analysis of 3D surfaces~\cite{su2015optimal}, etc. These applications were first proposed by Rubner et al.~\cite{Rubner2000}, under the name of the Earth Mover's Distance (EMD), which is in fact the Wasserstein metric of order one, $W_1$. More recent applications favor the $2^{\text{nd}}$ Wasserstein metric by way of Brenier's theorem~\cite{su2015optimal,de2011optimal}, which shows the uniqueness of the optimal solution for the squared-distance and its connection to differential geometry through which more efficient algorithms have been constructed.

This section is primarily concerned with the Wasserstein metric for discrete distributions in the Euclidean space. This allows us to focus on the concepts that are directly related to the practical calculation and estimation of this metric, and avoid the usage of measure theory without creating much ambiguity. The definition of the Wasserstein metric for general probability measures with more formal treatment of the probability theory is discussed in Appendix~\ref{APP_WM_GEN_PROB_MEASURE}. The interested reader is also referred to books by Villani~\cite{villani2008optimal}, surveys by Urbas~\cite{urbas1998mass}, and the more recent lectures by McCann and Guillen~\cite{mccann2011five} for further details.
\subsection{Preliminaries: Monge-Kantorovich transportation problem}
The Wasserstein metric is motivated by the classical optimal transportation problem first proposed by Monge in 1781~\cite{monge1781memoire}. The optimal transportation problem of Monge considers the most efficient transportation of ore from $n$ mines to $n$ factories, each of which produces and consumes one unit of ore respectively. These mines and factories form two finite point sets in the Euclidean space, which are denoted by $\cM$ and $\cF$. The cost of transporting one unit of ore from mine $x_i \in \cM$ to factory $y_j \in \cF$ is denoted by $c_{ij}$, which is chosen by Monge to be the Euclidean distance. The transport plan can be expressed in the form of an $n \times n$ matrix $\Gamma$, of which the element $\gamma_{ij}$ represent the amount of ore transported from $x_i$ to $y_j$. The central problem in optimal transportation is to find the transport plan that minimizes the total cost of transportation, which is the sum of the costs on $n \times n$ available routes between factories and mines, i.e., $\sum_{i=1}^{n} \sum_{j=1}^{n} \gamma_{ij} c_{ij}$. The original problem formulated by Monge requires transporting the ore in its entirety. Therefore, $\Gamma$ is constrained to be a permutation matrix, and can only take binary values of 0 or 1. Formally, the Monge transport problem can be expressed as 
\begin{equation}
\label{eq:monge_problem}
\begin{aligned}
& \underset{\Gamma}{\min}
& & \sum_{i=1}^{n} \sum_{j=1}^{n} \gamma_{ij} c_{ij} \\
& \text{subject to}
& & \sum_{i=1}^{n} \gamma_{ij} = \sum_{i=j}^{n} \gamma_{ij} = 1 \,,\quad \gamma_{ij} \in \{0,\, 1\} \, .
\end{aligned}
\end{equation}
The transport problem in Monge's formulation can be solved relatively efficiently by the Hungarian algorithm proposed by Kuhn in 1955~\cite{kuhn1955hungarian}.

In 1942, Kantorovich reformulated Monge's problem by relaxing the requirement that all the ore from a given mine goes to a single factory~\cite{kantorovich1958space, kantorovich2006problem}. As a result, the transport plan is modified from being a permutation matrix to a doubly stochastic matrix. With this, Monge's problem is written in the following form:

\begin{equation}
\label{eq:monge_kantorovich_problem}
\begin{aligned}
& \underset{\Gamma}{\min}
& & \sum_{i=1}^{n} \sum_{j=1}^{n} \gamma_{ij} c_{ij} \\
& \text{subject to}
& & \sum_{i=1}^{n} \gamma_{ij} = \sum_{i=j}^{n} \gamma_{ij} = 1 \,,\quad \gamma_{ij} \ge 0 \, .
\end{aligned}
\end{equation}
This relaxation eliminates the difficulties of Monge's formulation in terms of obtaining certain desirable mathematical properties, such as existence and uniqueness of the optimal transport plan. The Monge-Kantorovich transportation problem in Eq.~\ref{eq:monge_kantorovich_problem} can be solved as the solution to a general linear programming problem, although more efficient algorithms exists by exploiting special structures of the problem.

\subsection{\label{SSEC_WASSERSTEIN_METRIC}Wasserstein metric for discrete distributions}
The Monge-Kantorovich problem in Eq.~\ref{eq:monge_kantorovich_problem} not only gives the solution to the optimal transport problem, but also provides a way of quantifying the dissimilarity between the distributions of the mines and the factories by evaluating the optimal transport cost (normalized by total mass). The Wasserstein metric~\cite{dudley1976probabilities} follows directly from this observation.

Consider replacing the mines and factories by two general discrete distributions, whose probability mass functions are:
\begin{equation}
\label{eq:disc_dist}
	f(x) = \sum_{i = 1} ^{n} f_i \delta(x-x_i) \;,\qquad	g(y) = \sum_{j = 1} ^{n^{\prime}} g_j \delta(y-y_i) \, ,
\end{equation}
where $\sum_{i=1}^{n} f_i = \sum_{j=1}^{n^{\prime}} g_i = 1$ and $\delta(\cdot)$ denotes the Kronecker delta function:
\begin{equation}
    \delta (x)= 
\begin{cases}
    1	& \text{if } x = 0 \, ,\\
    0	& \text{if } \text{otherwise} \, .
\end{cases}
\end{equation}
The unit cost of transport between $x_i$ and $y_j$ is defined to be the $p^\text{th}$ power of the Euclidean distance, i.e.~$c_{ij}= |x_i - y_j|^p$. In addition, the ``mass" of probabilities $f_i$ and $g_j$ is no longer restricted to be unity and the possible outcomes of the distributions, i.e.~$x_i$ and $y_j$, are points in an Euclidean space whose dimension is not restricted.

The $p^\text{th}$ Wasserstein metric between discrete distributions $f$ and $g$ is defined to be the $p^\text{th}$ root of the optimal transport cost for the corresponding Monge-Kantorovich problem:
\begin{equation}
\label{eq:wasserstein_metric}
\begin{aligned}
& W_p(f,g) & & = \underset{\Gamma}{\min} \bigg(
\sum_{i=1}^{n} \sum_{j=1}^{n^{\prime}} \gamma_{ij} |x_i - y_j|^p \bigg)^{1/p} \\
& \text{subject to}
& & \sum_{j = 1} ^{n^{\prime}} \gamma_{ij} = f_i \,,\quad \sum_{i = 1} ^{n} \gamma_{ij} = g_j \,,\quad\gamma_{i,j} \ge 0 \, .
\end{aligned}
\end{equation}
The constraints in Eq.~\ref{eq:wasserstein_metric} ensure that the total mass transported from $x_i$ and the total mass transported to $y_j$
match $f_i$ and $g_j$, respectively. 

The Wasserstein metric for continuous distributions is presented in Appendix~\ref{APP_WM_GEN_PROB_MEASURE}. The estimation and calculation for the continuous case, for higher-dimensional problems, are discussed in the next section.

\subsection{\label{SSEC_WASSERSTEIN_MC}Non-parametric estimation of $W_p$ through empirical distributions}
Two major difficulties arise in obtaining the Wasserstein metric of two multivariate distributions of thermo-chemical states. One is that the multivariate distributions are not readily available from either experiments or simulations. Instead, a series of samples drawn from these distributions is provided. The other is that there is no easy way of calculating $W_p$ for continuous multivariate distributions, especially for those of high dimensions. To overcome these problems, a non-parametric estimation of $W_p$ is devised using empirical distributions. 

An empirical distribution is a random discrete distribution formed by a sequence of independent samples drawn from a given distribution of interest. Let $(\mathcal{X}_1,\, \ldots,\, \mathcal{X}_n)$ be a set of $n$ independent random samples obtained from a continuous multivariate distribution $f$. The empirical (or fine-grained~\cite{POPE_BOOK00}) distribution $\wh{f}$ is defined to be
\begin{equation}
\label{eq:emp_dist}
	\wh{f}_n(x) = \frac{1}{n}  \sum_{i = 1} ^{n} \delta(x-\mathcal{X}_i) \, ,
\end{equation}
which is a discrete distribution with equal weights. The empirical distribution is random as it depends on random samples, $\mathcal{X}_i$. The samples may be obtained from experimental measurements or generated from a given distribution using, for instance, an acceptance-rejection method~\cite{RUBINSTEIN_KROESE_BOOK2008}. Calculation of $W_p$ between two empirical distributions is identical to the method discussed in Sec.~\ref{SSEC_WASSERSTEIN_METRIC}. Its procedure and cost is independent of the dimensionality of the distributions. 

The empirical distribution converges to the original distribution. Most importantly, in the context of this study, is the convergence in the Wasserstein metric~\cite{bickel1981some}. More specifically, the Wasserstein metric between empirical and original distributions converges to zero in probability. Given the metric property of $W_p$:
\begin{equation}
\label{eq:wp_metric_prop}
|W_p(\wh{f}_n, \wh{g}_{n^\prime}) - W_p(f, g)| \le W_p(\wh{f}_n, f) + W_p(\wh{g}_{n^\prime}, g) \, ,
\end{equation}
such convergence ensures that the Wasserstein metric between two empirical distributions also converges in probability to that of the actual distribution. In addition, the mean rates of convergence for empirical distributions have also been established~\cite{horowitz1994mean,fournier2015rate}, giving the upper bound on the expectation, $E\big(W_p^p(\wh{f}_n,\, f)\big),$ as $n$ increases. For the similar reason, the convergence rate for the non-parametric estimation follows. The exact rates depend on the dimensionality and regularity conditions of the distributions. For details on the convergence rate for more general cases, the interested reader is referred to the work by Fournier and Guillin~\cite{fournier2015rate}. 

In the case of $d$-dimensional distributions that have sufficiently many moments, we have:
\begin{equation}
	E\left(W_2^2(\wh{f}_n, f)\right) \le C \times
\begin{cases}
    n^{-1/2}		& \text{if } d < 4 \, ,\\
    n^{-1/2}	\log(1+n) & \text{if } d = 4 \, , \\
     n^{-2/d} & \text{if } d > 4 \, .
\end{cases}
\end{equation}
where the value of $C$ depends on the distribution $f$ and is independent of $n$. By combining this result with Eq.~\ref{eq:wp_metric_prop}, we obtain the following convergence rate for the non-parametric estimation of $W_2$:
\begin{equation}
\label{e:rate_of_convergence_w2}
	E\left(W_2(\wh{f}_n, \wh{g}_{n^\prime}) - W_2(f, g)\right)^2 \le C \times
\begin{cases}
    n_{*}^{-1/2}		& \text{if } d < 4 \, ,\\
    n_{*}^{-1/2}	\log(1+n_{*}) & \text{if } d = 4 \, , \\
    n_{*}^{-2/d} & \text{if } d > 4 \, ,
\end{cases}
\end{equation}
where $n_{*} = \min(n,\, n^\prime)$. With this, all necessary results that ensure the soundness of using $W_p(\wh{f}_n, \wh{g}_{n^\prime})$ as an estimator for $W_p(f, g)$ are presented. 

In addition to the rate of convergence , statistical inference and further quantification of uncertainty for the non-parametric estimation of $W_p$ can also be performed. In particular, the magnitude of the uncertainty in Eq.~\ref{e:rate_of_convergence_w2} and the corresponding confidence interval can be estimated via the method of bootstrap~\cite{efron1979bootstrap,chernick2011bootstrap}. In particular the m-out-n bootstrap~\cite{bickel1996m} is shown to be consistent for the Wasserstein metric~\cite{sommerfeld2016inference,dumbgen1993nondifferentiable}.

\subsection{\label{SSEC_SML}Statistically most likely reconstruction of distributions}
So far, the evaluation of the Wasserstein metric using sample data as empirical distribution function has been described. However, the usage of the Wasserstein metric is not limited to results reported in such fashion. In the following, the computation of the Wasserstein metric from statistical results will be discussed. This versatility is important to the metric, given the fact that the conventional practice of reporting only the statistics, predominantly first and second moments, is still the prevailing one. 

The procedure of applying the Wasserstein metric to statistical results is to first reconstruct the multivariate distribution from statistical results. An empirical distribution is then sampled to compute the Wasserstein metric following the method described in Sec.~\ref{SSEC_CALC_PROC}. The reconstruction of a continuous PDF from a set of known statistical models can be performed using the concept of the statistically most likely distribution (SMLD)~\cite{POPE_JNT79,IHME_PITSCH_PRT1_CF_2008,Coclite2015}. The SMLD of a $d$-dimensional random variable is defined to be the distribution that maximizes the relative entropy, given a prior distribution $g(\mathbf{x})$, under a given set of constraints:
\begin{equation}
\label{eq:SMLD}
\begin{aligned}
& f_\text{SML}(\mathbf{x}) 
& & = \underset{f(\cdot)}{\operatorname{argmax}} \int_{\mathcal{R}^d} f(\mathbf{x}) \ln \left(\frac{f(\mathbf{x})}{g(\mathbf{x})}\right) d\mathbf{x} \\
& \text{subject to} 
& & \int_{\mathcal{R}^d} f(\mathbf{x}) d\mathbf{x} = 1\,, \\
& & & \int_{\mathcal{R}^d} \mathbf{T}(\mathbf{x}) f(\mathbf{x}) d\mathbf{x} = \overline{\mathbf{t}} \,. \\
\end{aligned}
\end{equation}
Here, $\mathbf{T}(\mathbf{x})$ is the set of statistical moments that are selected as constraints, and $\overline{\mathbf{t}}$ is the vector of corresponding values obtained from the data. The type of the so obtained distribution is dictated by the form of the constraints~\cite{lisman1972note}. For instance, if the multivariate distribution is constructed using only the first and second moments with a uniform prior, a multivariate normal distribution is obtained. In addition, if only the marginal second moments are given while the cross moments are not, the obtained multivariate normal distributions are uncorrelated. More specifically, if 
\begin{equation}
\begin{aligned}
g(\mathbf{x}) &\equiv 1 \, , \\
\mathbf{T}(\mathbf{x}) & = [\mathbf{x}_1,\mathbf{x}_2,\ldots,\mathbf{x}_d, \mathbf{x}_1^2, \mathbf{x}_2^2,\ldots, \mathbf{x}_d^2]\,,
\end{aligned}
\end{equation}
the SMLD becomes
\begin{equation}
f_\text{SML}(\mathbf{x}) = \prod_{k = 1}^{d} \frac{1}{\sqrt{2\pi \boldsymbol{\sigma}_k^2}} \exp{ \left\{ -\f{(\mathbf{x}_k - \boldsymbol{\mu}_k)^2}{ 2 \boldsymbol{\sigma}_k^2} \right\} } \, ,
\end{equation}
where $\boldsymbol{\mu}_k = \overline{\mathbf{x}}_k$ and $\boldsymbol{\sigma}_k^2 = \overline {\mathbf{x}^2_k} - \overline {\mathbf{x}_k}^2$.

After obtaining the SMLDs, the sets of samples can be drawn from them and the Wasserstein metric can be directly calculated. In addition, the metric property implies the following inequality
\begin{equation}
	|W_p(f_\text{SML},g_\text{SML}) - W_p(f,g)| \le W_p(f_\text{SML},f) +  W_p(g_\text{SML},g) \, .
\end{equation}
In other words, the difference in the Wasserstein metrics of the actual distributions, $f$ and $g$, and that of the reconstructed counterparts, $f_\text{SML}$ and $g_\text{SML}$, are bounded by the error of the SMLD reconstruction, which themselves can be quantified by the Wasserstein metrics. In the current study, the set of constraints are limited to the marginal first and second moments, which are typically reported in the literature. More accurate reconstruction can be obtained by including higher-order and cross moments. In addition, statistics of other forms may also be considered. For instance, $\beta$-distributions can be recovered for constraints in the form of $\overline{\ln(x)}$ and $\overline{\ln(1-x)}$, which is potentially more appropriate for conservative scalars such as mixture fraction.

\subsection{\label{SSEC_CALC_PROC}Calculation procedure}
\begin{figure}[!htb!]
\begin{center}
 \subfigure[\label{FIG_W2_DIST} PDF Sampling.]{\includegraphics[width = 0.45\textwidth]{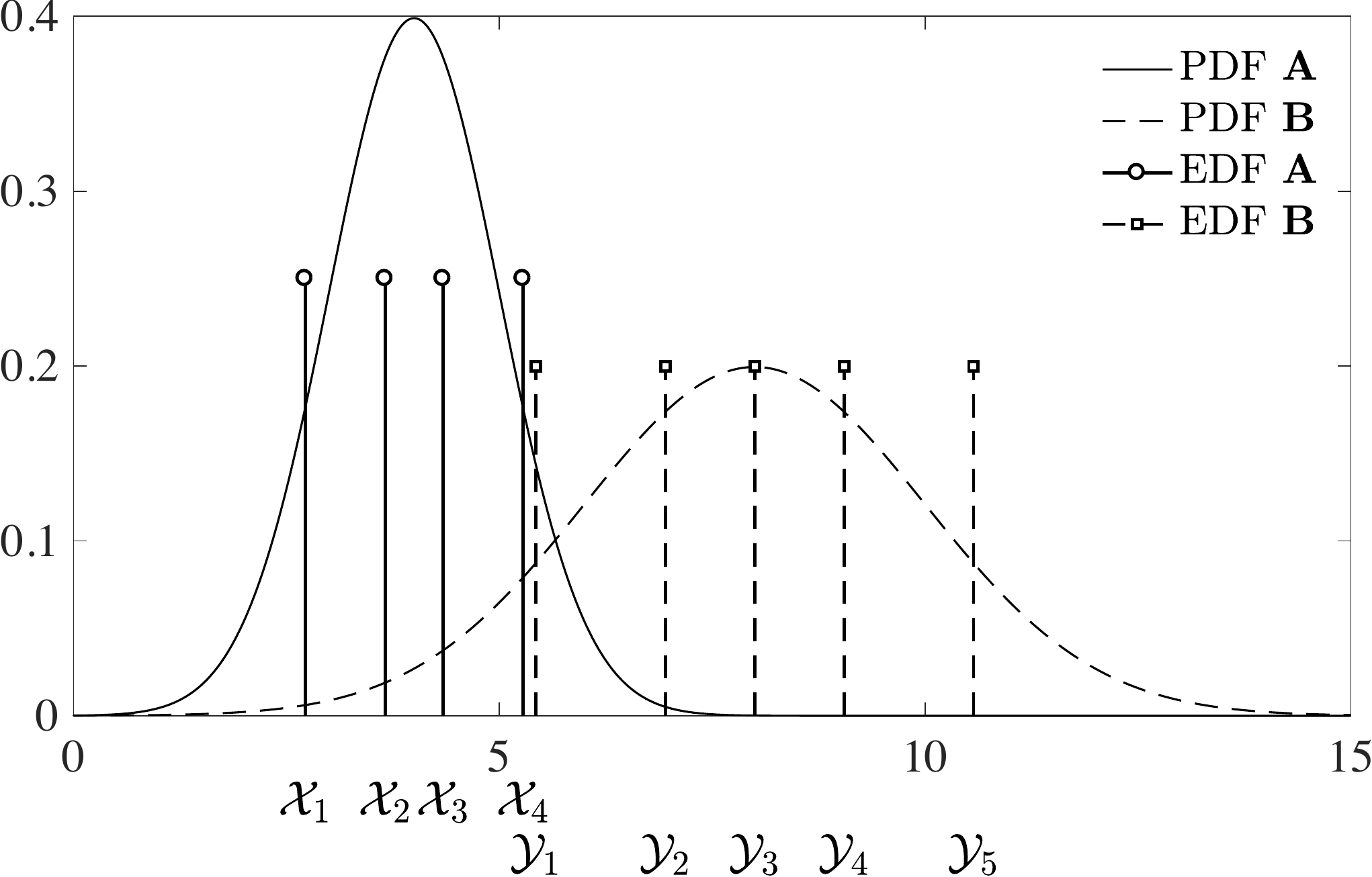}}\\
 \subfigure[\label{FIG_W2_FLOW}Transport matrix $\Gamma$.]{
    \begin{tabular}{|c||*{5}c|}
    \hline
    \diagbox{\small$\mathcal{X}$}{\small$\mathcal{Y}$} & 5.43 & 6.95 & 8.00 & 9.05 & 10.57\\\hline\hline
    2.72 & 0.15 & 0.10 &      &      & \\
    3.66 &      & 0.05 & 0.20 &      & \\
    4.34 &      & 0.05 &      & 0.20 & \\
    5.28 & 0.05 &      &      &      & 0.20\\\hline
    \multicolumn{6}{c}{}\\[-2ex]
    \end{tabular}
    \vspace*{4cm}
}
 \caption{\label{FIG_W2}Illustrative example of computing the Wasserstein matrix from two distribution functions: (a) The continuous PDFs are first sampled by empirical distribution functions (EDF, shown by discrete peaks, which are here down-sampled for illustrated purposes). The Wasserstein metric is then evaluated from the samples ${\mathcal{X}}_{i=1,\ldots,4}$ and ${\mathcal{Y}}_{i=1,\ldots,5}$ with (b) coefficient of transport matrix $\Gamma$, obtained from the solution to the minimization problem of Eq.~\ref{eq:wasserstein_metric}.}
\end{center}
\end{figure}

In the present study, the $2^{\text{nd}}$ Wasserstein metric is used. The computation of the metric can be realized by any general-purpose linear programming tool. For the present analysis, the program by Pele and Werman~\cite{Pele2009} is used. It calculates the Wasserstein metric as a flow-min-cost problem (a special case of linear programming) using the successive shortest path algorithm~\cite{ahuja1988network}. A pseudocode of the corresponding algorithm is given in Alg.~\ref{alg:Wasserstein_metric}, and Fig.~\ref{FIG_W2} provides an illustrative example for the evaluation of the Wasserstein metric. The source code for the evaluation of the Wasserstein metric is provided in Sec.~\ref{APP_SRC_CODE}. 

In this context, it is important to note that the input data to the Wasserstein metric are normalized to enable a direct comparison and enable a physical interpretation of the results. A natural choice is to normalize each sample-space variable by its respective standard deviation that is computed from the reference data set (for instance, the experimental measurements).
\begin{algorithm}[!hbt!]
	\SetKwInput{Input}{Input}
    \SetKwInput{Output}{Output}
	\SetKwInput{Preprocessing}{Preprocessing}
	\SetKwInput{Processing}{Processing}
	\Input{Two sets of $d$-dimensional data representing empirical distribution function: $\mathcal{X}$, $\mathcal{Y}$, with lengths $n$ and $n^{\prime}$ from scatter data or sampled from continuous distribution function}
    \Preprocessing{
	Normalize $\mathcal{X}$ and $\mathcal{Y}$ by standard deviation of one data set, $\boldsymbol{\sigma}_\mathbf{x}$
    }
	\For{i = $1: \, n$}{
		\For{j = $1: \,n^{\prime}$}{		
			Evaluate pair-wise distance matrix $c_{i,j} =\sum_{k=1}^{d} 
			\left(\mathcal{X}_{k,i}-\mathcal{Y}_{k,j}\right)^2$; }}
	Compute Wasserstein metric and transport matrix as solution to minimization problem of Eq.~\ref{eq:wasserstein_metric} using the shortest path algorithm by Pele \& Werman~\cite{Pele2009} with input $c_{i,j}$\\
	\Output{Wasserstein metric: $W_2$}
 	\caption{Pseudocode for evaluating the Wasserstein metric.}
	\label{alg:Wasserstein_metric}
\end{algorithm}

Suppose we have two sets of data with sample sizes of $n$ and $n^{\prime}$, respectively. Each sample represents a point in the thermo-chemical (sub)-space, e.g.\ $\mathbf{x} = [ Z,T,Y_{\ce{H2O}},\ldots ]$. The empirical distribution can be constructed  from each data set following Eq. \ref{eq:emp_dist}, where $f_i = 1/{n}$ and $g_i = 1/{n^{\prime}}$. The Wasserstein metric is then computed following the definition in Eq.~\ref{eq:wasserstein_metric}. The worst time complexity of the algorithm is $\mathcal{O} \left((n+n^{\prime})^3 \log(n + n^{\prime})\right)$. Note that the dimension of the thermo-chemical (sub)-space affects only the pair-wise distance between data points but not the definition or calculation of the metric.

\subsection{Remarks on the interpretation and usage of the Wasserstein metric $W_p$}

The Wasserstein metric is a natural extension of the Euclidean distance to statistical distributions. It enables the comparison between two multi-dimensional distributions via a single metric while taking all information presented by the distributions into consideration. The definition of $W_p$ in Eq.~\ref{eq:wasserstein_metric} can be viewed as the weighted average of the pair-wise distances between samples of the two distributions. In the case of one-dimensional distributions, the obtained value of the metric shares the same unit as the sample data. For instance, if two distributions of temperature are considered, the corresponding $W_p$ in unit of kelvin can be interpreted as the average difference between the values of temperature from the two distributions. In the case of multi-dimensional distributions, each dimension is normalized before pair-wise distances are calculated. The choice of the normalization method is application-specific. In this study, the marginal standard deviation is chosen. The so obtained $W_p$ represents the averaged difference, proportional to the marginal standard deviations, between samples from the two distributions. As such, a value of $W_p = 0.5$ can be interpreted as a difference of simulation and experimental data at the level of $0.5$ standard deviation. Although not considered in this study, additional turbulent-relevant information, such as space-time correlation, can be factored in via the extension of the phase space for the PDFs as performed by Muskulus and Verduyn-Lunel~\cite{muskulus2011wasserstein}.

Comparable samples need to be drawn from numerical simulation and experimental data to ensure a consistent comparison between the two distributions. This can be achieved by matching the sampling locations and frequencies between the two sources of data. Furthermore, the experimental uncertainty may also be factored in by either the convolution of simulation data with error distributions or the Bayes deconvolution of the experimental data~\cite{doi:10.1093/biomet/asv068,laird1978nonparametric}.

\section{\label{SEC_CONFIG}Configuration}
In the present study, the utility of the Wasserstein metric is evaluated in an application to simulation results of a piloted turbulent jet flame with inhomogeneous inlet. The experimental configuration is described in the next section, the computational setup is summarized in Sec.~\ref{SUBSEC_CONFIG_COMP}, and statistical simulation results that are utilized in evaluating Wasserstein metric are presented in Sec.~\ref{SUBSEC_CONFIG_RESULTS}.
\subsection{\label{SSEC_CONFIG_EXP}Experimental setup}
\begin{figure}[!htb!]
\centering
\includegraphics[width=0.2\textwidth]{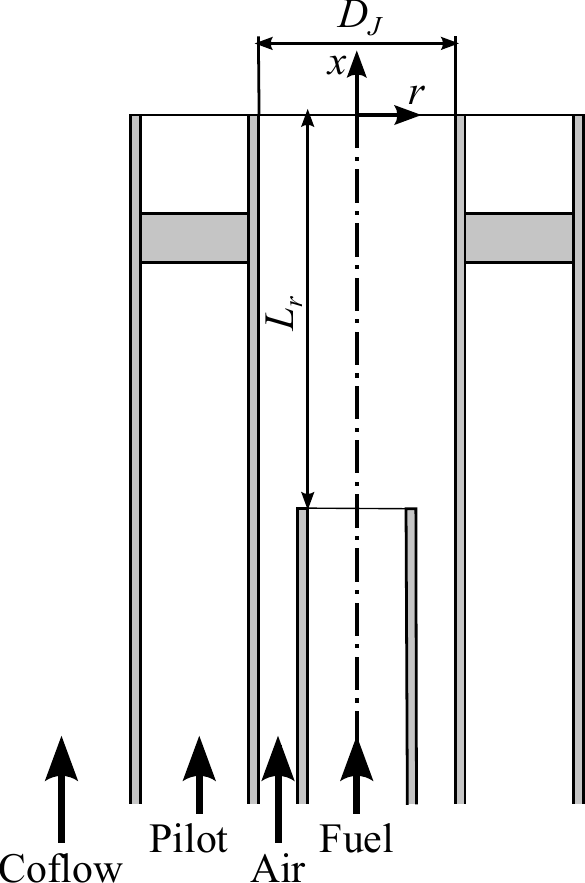}
\caption{Schematic of the piloted turbulent burner with inhomogeneous inlets. \label{fig:burner}
}
\end{figure}

The piloted turbulent jet flame with inhomogeneous inlets considered in this work was experimentally investigated by Meares et al.~\cite{MEARES_MASRI_CF2014,MEARES_PRASAD_MAGNOTTI_BARLOW_MASRI_PCI2015} and Barlow et al.~\cite{BARLOW_MEARES_MAGNOTTI_CUTCHER_MASRI_CF2015}. The burner is schematically shown in Fig.~\ref{fig:burner}. The reactants are supplied through an injector that consists of three tubes; the inner fuel-supply tube with a diameter of 4 mm is recessed by a length $L_r$ with respect to the burner exit plane. This inner tube is placed inside an outer tube (with a diameter of $D_J=7.5$ mm) of ambient air; depending on the recess height, varying levels of mixture inhomogeneity can be achieved. The flame is stabilized by a pilot stream, exiting through an annulus with outer diameter of 18 mm. The burner is placed inside a wind tunnel, providing co-flowing air at a bulk velocity of 15 m/s.

The present study considers the operating condition \emph{FJ-5GP-Lr75-57}. The fuel of methane is supplied through the inner tube (\emph{FJ}). The pilot flame is a gas mixture consisting of five components (\emph{5GP}: $\ce{C_2H_2}$, $\ce{H_2}$, $\ce{CO_2}$, $\ce{N_2}$, and air to match the product gas composition and equilibrium temperature of stoichiometric \ce{CH4}/air mixture). The bulk velocity of the unburned pilot mixture is $3.72$ m/s. The inner tube of the fuel stream is recessed from the jet exit plane by $L_r=75\,\text{mm}$, with the bulk jet velocity is set to 57 m/s (\emph{Lr75-57}), corresponding to $50\%$ of the experimentally measured blow-off velocity. The recess results in a partially premixed reactant-gas mixture, which is relevant to modern gas turbine applications~\cite{moin2006large,esclapez2017fuel}.

\subsection{\label{SUBSEC_CONFIG_COMP}Computational setup and mathematical model}
The computational domain is discretized using a three-dimensional structured mesh in cylindrical coordinates, and includes the upstream portion of the burner to represent the partial mixing of reactants and flame stabilization. The computational domain extends to $20D_J$ in axial direction and $15D_J$ in radial direction, and is discretized using 1.6 million control volumes. Inflow conditions for the fuel/air jet, the pilot, and the coflowing air stream are obtained from separate simulations. The pilot flame is treated by prescribing the scalar profile from the corresponding chemistry table, with the mixture stoichiometry, temperature, and mass flow rate representing the experimental setting. An improved flame stability is experimentally observed with the inhomogeneous inlets condition. This can be attributed to the upstream premixed combustion of the near-stoichiometric fluid in the jet reacting with the hot pilot. Local extinction and re-ignition was found to be not relevant under these operating conditions~\cite{MEARES_PRASAD_MAGNOTTI_BARLOW_MASRI_PCI2015}.

To model the turbulent reacting flow-field, a flamelet/progress variable (FPV) model is employed~\cite{CHUCK_JFM04,IHME_PCI04}, in which the thermo-chemical quantities are expressed in terms of a reaction-transport manifold that is constructed from the solution of steady-state non-premixed flamelet equations~\cite{PETERS_PECS84}. This model requires the solution of transport equations for the filtered mixture fraction, residual mixture fraction variance, and filtered reaction progress variable. These modeled equations take the following form:
\begin{subeqnarray}
 \label{EQ_TE}
 \slabel{EQ_TE_ZMEAN}
  \ol{\rho} \wt{\cD}_t\wt{Z} &=& \nabla \cdot (\ol{\rho}\wt{\alpha}\nabla \wt{Z}) +
                                 \nabla \cdot\tauvec^{\rm{res}}_{\wt{Z}}\;,\\
 \slabel{EQ_TE_ZVAR}
  \ol{\rho} \wt{\cD}_t\wt{Z''^2} &=& \nabla \cdot (\ol{\rho}\wt{\alpha}\nabla \wt{Z''^2}) +
                                 \nabla \cdot\tauvec^{\rm{res}}_{\wt{Z''^2}} - 
                                 2\ol{\rho}\wt{\uvec''Z''}\cdot\nabla\wt{Z} - 
                                 \ol{\rho}\wt{\chi}^{\rm{res}}_Z\;,\\
 \slabel{EQ_TE_CMEAN}
  \ol{\rho} \wt{\cD}_t\wt{C} &=& \nabla \cdot (\ol{\rho}\wt{\alpha}\nabla \wt{C}) +
                                 \nabla \cdot\tauvec^{\rm{res}}_{\wt{C}} + \ol{\rho}\wt{\dot{\omega}}_C\;,
\end{subeqnarray}
in which the turbulent fluxes are modeled by a gradient transport assumption~\cite{POINSOTBOOK01}, and the residual scalar dissipation rate $\wt{\chi}^{\rm{res}}_Z$ is evaluated using spectral arguments~\cite{IHME_PITSCH_PRT2_CF_2008}. With the solution of Eqs.~\ref{EQ_TE}, all thermo-chemical quantities are then expressed in terms of $\wt{Z},\wt{Z''^2},$ and $\wt{C}$, and a presumed PDF-closure is used to model the turbulence-chemistry interaction. For this, the marginal PDF for mixture fraction is described by a presumed $\beta$-PDF, and the conditional PDF of the reaction progress variable is modeled as a Dirac-delta function. 

A recently performed analysis of the model compliance showed that the FPV-approach only provides an incomplete description of the interaction between the partially premixed mixture and the hot pilot and discrepancies in the prediction of carbon monoxide~\cite{WU_IHME_FUEL2016}. Therefore, the present simulation is intended for the purpose of demonstrating the merit of employing the Wasserstein metric as a quantitative validation measure and to identify discrepancies of the model through direct comparisons against experiments. Extended flamelet models have been developed to describe the complex flame topology and turbulence-chemistry interaction appearing in this configuration~\cite{CHEN_IHME_CF2013,IHME_SEE_PCI33,WU_SEE_WANG_IHME_CF2015,wang2017regularized}, and the performance of other models will be examined in Sec.~\ref{SEC_TNF}.
\subsection{\label{SUBSEC_CONFIG_RESULTS}Statistical results and scatter data}
Before we examine the Wasserstein metric, this section summarizes statistical results between simulations and experiments. The simulated data presented in this section is not expected to replicate the experimental results perfectly. Instead, the data serves the purpose to determine where in the flame and for which species the model behaves well, as well as regions in which the model could be improved. Quantitative results using the Wasserstein validation metric will be presented in Sec.~\ref{SEC_RESULTS}, and these results should match the data interpretation developed in this section.

\begin{figure}[!htb!]
\centering
\includegraphics[width=\textwidth]{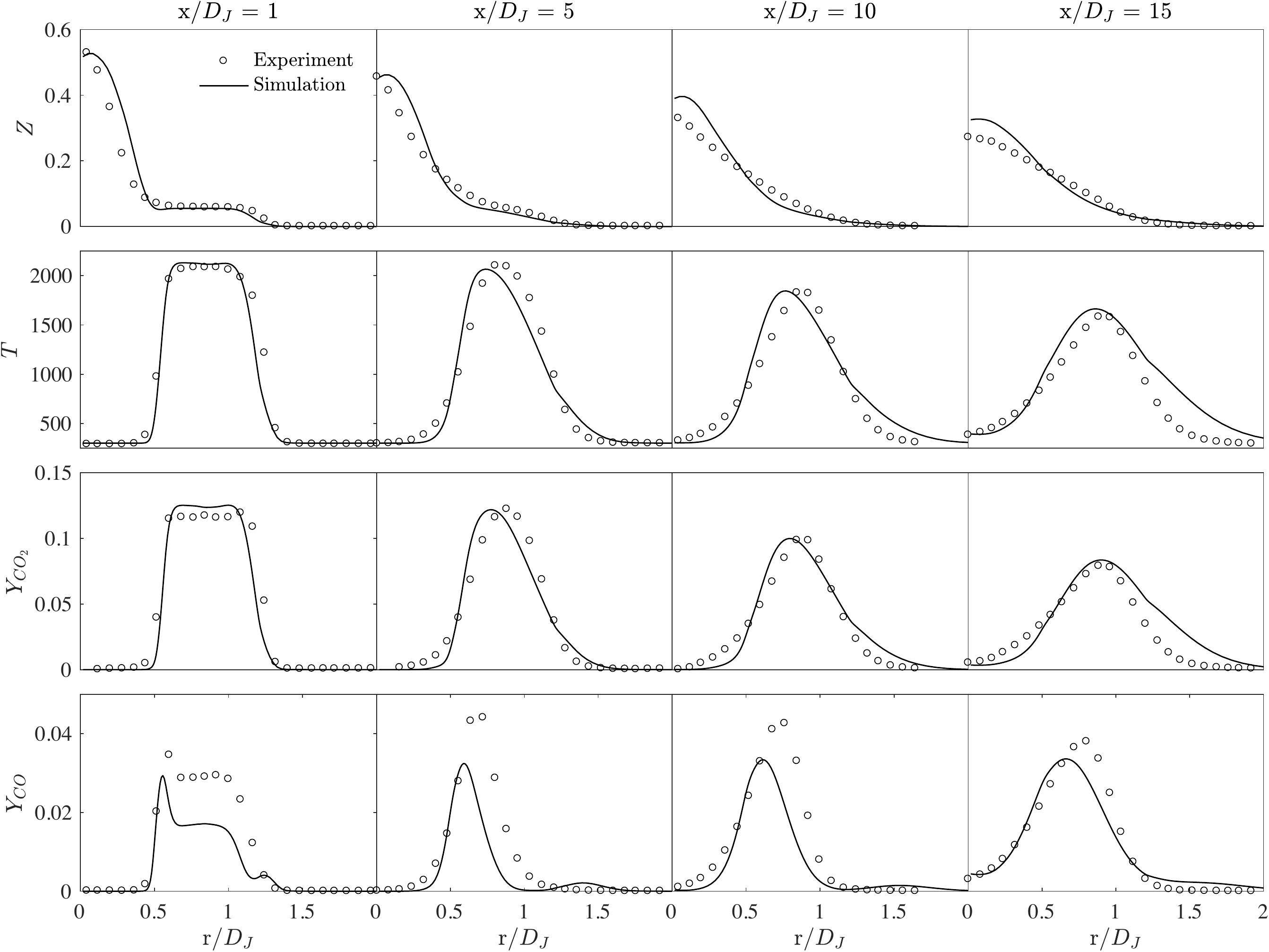}
\caption{Comparison of radial profiles between experimental measurements and simulations at $x/D_J = \{1,5,10,15\}$. Variables considered include mixture fraction, temperature, and species mass fractions of \ce{CO_{2}} and \ce{CO}. \label{fig:FPV_radial} 
}
\end{figure}
Comparisons between radial profiles of experimental data and simulations are provided in Fig.~\ref{fig:FPV_radial}. The comparisons are made at four distinct axial locations ($x/D_J = \{1, 5, 10, 15\}$), with four scalar quantities, which include mixture fraction ($Z$), temperature ($T$), and species mass fractions of \ce{CO2} and \ce{CO}. The solid lines represent simulation results, while the symbols correspond to data collected from experimental data. In the following, these four scalars are used as quantities of interest for the Wasserstein metric to embody the accuracy of the simulations in modeling mixing, heat release, fuel conversion, and emissions. Radial profiles for mixture fraction and temperature are in overall good agreement with measurements. Discrepancies are largely confined to regions in the jet core and shear-layer, where simulations underpredict scalar mixing. A shift in the peak location for temperature and \ce{CO2} mass fraction profiles at the intermediate axial locations, $x/D_J = \{5,10\}$ is apparent, which can be related to discrepancies in the mixing profile. Results for mean CO-profiles are presented in the last row of Fig.~\ref{fig:FPV_radial}, showing that the simulation underpredicts this intermediate product, which can be attributed to shortcomings of the FPV-combustion model~\cite{WU_IHME_FUEL2016}.

\begin{figure}[!htb!]
\centering
 \includegraphics[width= \textwidth]{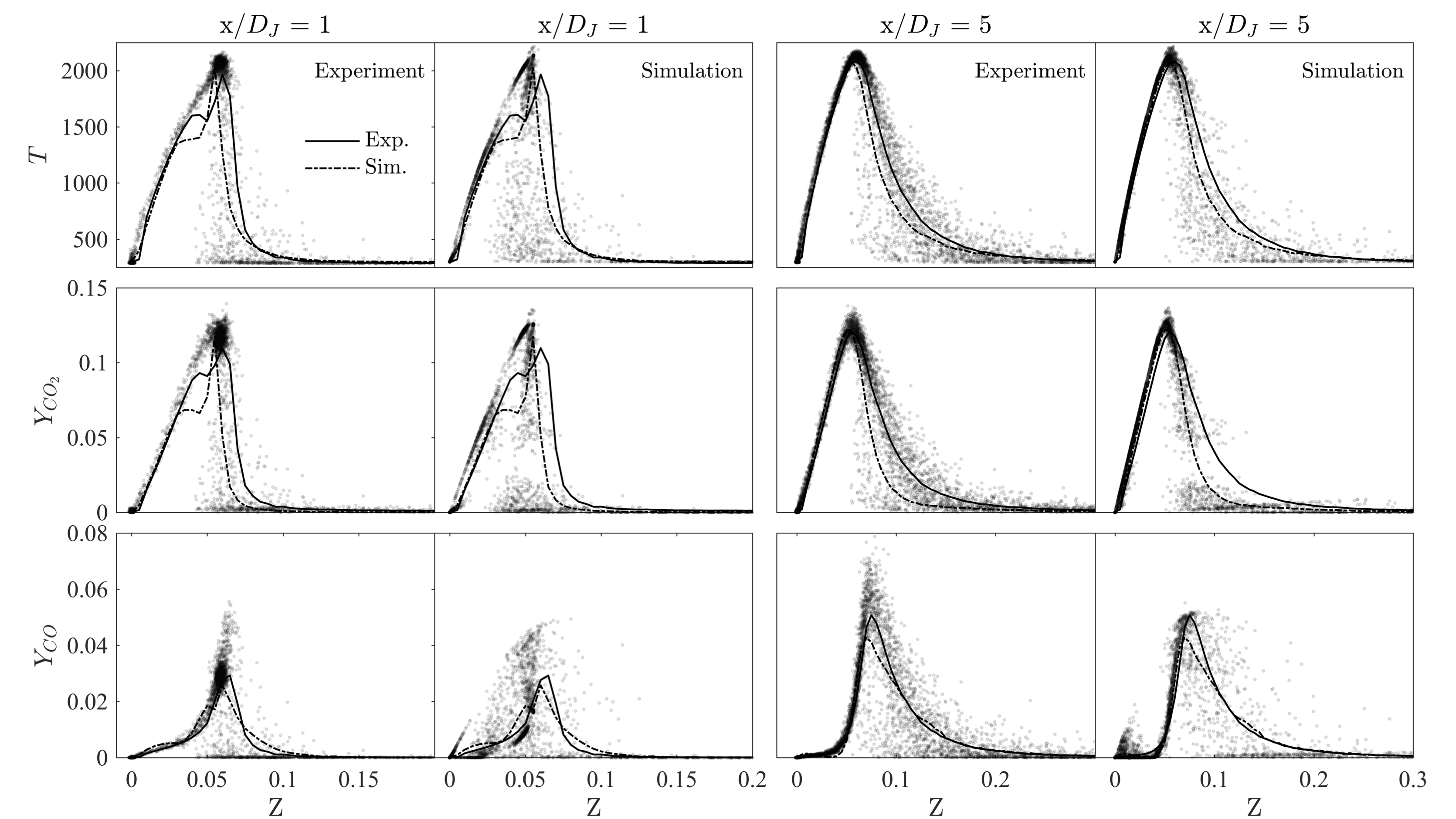}
\caption{\label{fig:FPV_scatter}Qualitative comparison of scatter profiles between measurements and simulations at $x/D_J = \{1, 5\}$. Variables including temperature, and species mass fractions of \ce{CO_{2}} and \ce{CO} are plotted against mixture fraction. Conditional mean results are laid over the scatter data.}
\end{figure}
Scatter data and mixture-fraction conditioned data from experiments and simulations are shown in Fig.~\ref{fig:FPV_scatter}. This data is sampled at a subset of the axial locations, while using $Z$, $T$, and mass fractions of \ce{CO2} and \ce{CO} as the same four quantities of interest. Scatter data are frequently examined to assess the agreement of the thermo-chemical state space that is accessed by the model and experiment. While this direct comparison provides insight about shifts in the composition profiles, as seen for mass fractions of CO and \ce{CO2}, such comparisons are mostly of qualitative nature. By utilizing the Wasserstein metric, these scatter data and statistical results will be used in the next section to obtain an objective measure for the quantitative assessment of the agreement between measurements and simulations.

\section{\label{SEC_RESULTS} Results for application of Wasserstein metric}
To introduce the Wasserstein metric as a quantitative validation tool, in the following we consider two test cases. The first test case, presented in Sec.~\ref{SSEC_CONSERVED}, focuses on the analysis of a single-scalar experimental results in which mixture fraction data is considered at individual points in the flame. 
Previous work has shown that the mixture fraction can reasonably be approximated by a $\beta$-distribution~\cite{Cook1994}, and this test case examines this premise by applying the Wasserstein metric to experimental data and modeled $\beta$-distribution that is obtained from a maximum likelihood estimation (MLE) of the measurements. This one-dimensional test case is intended to present the capabilities of the Wasserstein metric in a simplified context.

The second test, presented in Sec.~\ref{SSEC_MULTI}, considers the quantitative validation of LES modeling results against experimental data. For this, the Wasserstein metric will be employed to incorporate multiple thermo-chemical quantities, including $Z$, $T$, $Y_{\ce{CO2}}$, and $Y_{\ce{CO}}$, thereby contracting information about the model accuracy for predicting mixing, fuel conversion, and emissions into a single validation measure. Several locations in the flame will be considered to evaluate potential model deficiencies, demonstrating the merit of the Wasserstein metric as multidimensional validation tool.

In these test cases, scatter data and empirical distributions that are reconstructed from statistical moment information, using the method presented in Sec.~\ref{SSEC_SML}, are considered to examine the accuracy of both methods.
\subsection{\label{SSEC_CONSERVED}Conserved scalar results}
Previous investigations have shown that the evolution of conserved scalars in two-stream systems can be approximated by a $\beta$-distribution~\cite{Cook1994}, and a Dirichlet-distribution as a multivariate generalization of the $\beta$-distribution provides a description of turbulent scalar mixing in multistream flows~\cite{CHEN_IHME_CF2013}. This section examines the accuracy of representing the conserved scalar by a presumed PDF using the Wasserstein metric as a quantitative metric. To simplify the analysis, this study focuses on data collected at the axial locations introduced previously ($x/D_J = \{1, 5, 10, 15\}$) and a set of four radial positions located at $r/D_J = \{0, 0.5, 0.85, 1.2\}$. The axial positions are spaced uniformly, whereas the radial positions represent key locations in the burner geometry. Specifically, these four radial locations correspond approximately to the center of the fuel stream, the outer edge of the air stream, the midpoint of the pilot, and the outer edge of the pilot, respectively. Although these sixteen measurement locations were chosen for the analysis, it is possible to apply the Wasserstein metric at any location in the flame and generate similar results.  

\begin{figure}[!htb!]
\begin{center}
 \includegraphics[width = \textwidth]{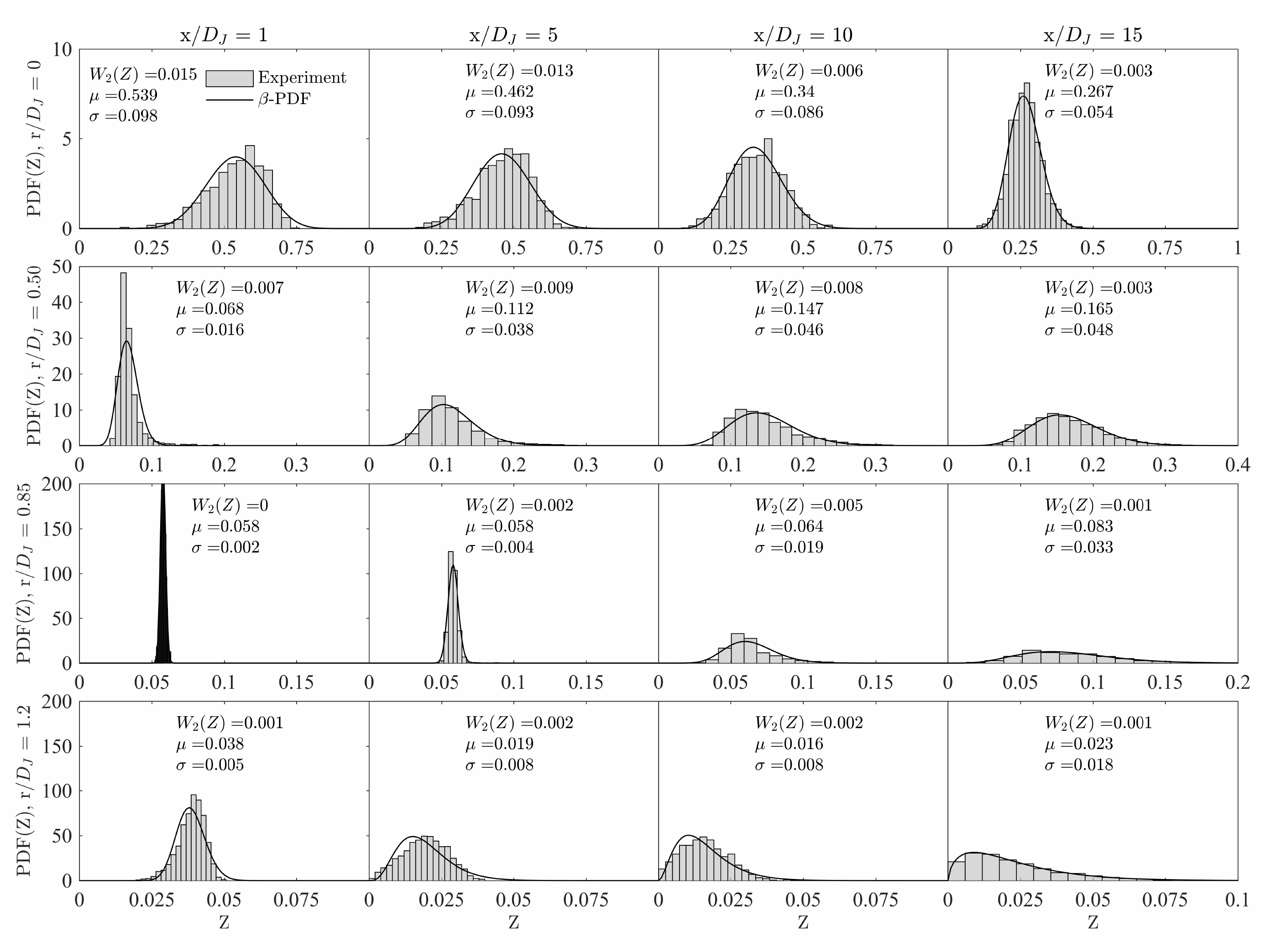}
 \caption{\label{FIG_B_MATRIX} Distributions of mixture fraction from measurements and presumed $\beta$-distribution at different locations in the piloted partially premixed jet flame with inhomogeneous inlets. The Wasserstein metric is calculated for each location, showing that modeling performance improves with increasing axial and radial distance. For reference, values for mean ($\mu$) and standard deviation ($\sigma$), computed from the experimental data, are provided. } 
\end{center}
\end{figure}

PDFs for mixture fraction, represented by the histograms in Fig.~\ref{FIG_B_MATRIX}, provide the experimental distribution for all of the points of interest. Superimposed over each plot is a maximum likelihood estimation of this data using the $\beta$-distribution~\cite{rice2006mathematical}, whose probability distribution function reads
\begin{equation}
  f(x;a,b) = \frac{\Gamma(a+b)}{\Gamma(a)\Gamma(b)} x^{a-1}(1-x)^{(b-1)}\;,
\end{equation}
where $\Gamma(\cdot)$ is the gamma function. Suppose that there are $n$ independent samples drawn from a $\beta$-distribution, whose values are $x_1,\, x_2,\, \ldots, \, x_n$, the method of maximum likelihood estimates the parameter $a$ and $b$ by finding the arguments of the maxima for the logarithmic likelihood function,
\begin{equation}
	\wh{a},\widehat{b} = \text{arg}\,\max\limits_{a,b} \sum_{i=1}^{n} \log \bigg( f(x_i;a,b) \bigg) \, .
\end{equation}
Although the best-fit $\beta$-distribution provides a reasonable approximation for the data, there still are noticeable differences between the experimental data and the MLE-fit. These differences are quantitatively expressed through the Wasserstein metric, and the computed values are reported at the top of each histogram in Fig.~\ref{FIG_B_MATRIX}.

The quantitative evaluation of the Wasserstein metric shows that largest deviations between experimental and presumed distributions occur within the fuel jet near the nozzle exit. These quantitative results are corroborated with an interpretation of the histograms. The largest deviations between the MLE $\beta$-distribution and measurements occur near the nozzle inlet, corresponding to regions of high turbulence and strong mixing. Note that the magnitude of the Wasserstein metric, $W_2(Z)$, is provided in natural units of mixture fraction, which is bounded to the interval $[0,1]$. As such, the metric provides an integral physical interpretation of the differences between both distributions.

\begin{figure}[!htb!]
\begin{center}
 \includegraphics[width = \textwidth]{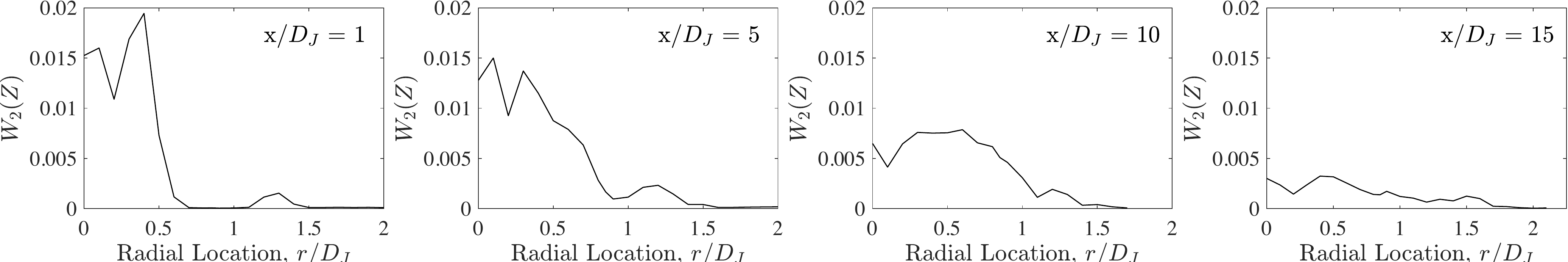}
 \caption{\label{FIG_B_RADIAL}Computed Wasserstein metric for mixture fraction between measurements and presumed $\beta$-distribution obtained from MLE, along radial directions at $x/D_J = \{1, 5, 10, 15\}$. }
\end{center}
\end{figure}

In the next step, we employ the Wasserstein metric to evaluate the accuracy of the presumed $\beta$-distribution along several radial profiles. For this, a total of 86 uniformly spaced points are considered along the four axial measurement stations used previously. At each point, a $\beta$-distribution is constructed from the measurements using the maximum likelihood estimate, from which calculations of the Wasserstein metric are performed subsequently. Results are presented in Fig.~\ref{FIG_B_RADIAL}, and they show that the agreement of the $\beta$-distribution with the experimentally determined PDFs improves with increasing axial and radial distance. This observation corroborates the findings from the point-wise analysis in Fig.~\ref{FIG_B_MATRIX}, and agrees with physical expectation that with increasing downstream distance the mixture composition approaches a homogeneous state. The low absolute error values of $W_2 < 0.02$ compared to the [0,1] range of mixture fraction shows that, overall, the $\beta$-distribution provides an adequate representation of the experimentally determined mixture fraction data.

While this test case affirms that the mixture fraction PDF follows a $\beta$-distribution, it also demonstrates three key traits of the Wasserstein metric. First, the quantitative nature of the Wasserstein metric allowed for direct comparisons between several distributions, simultaneously. For example, it is much easier to compare the four distributions at $x/D_J = 1$ using the Wasserstein metric results, as opposed to analyzing their differences in Fig.~\ref{FIG_B_MATRIX} directly. The Wasserstein metric also provides a comparison in distribution space. It therefore contains information about all moments, and is not limited to low-order moments such as mean and root-mean square quantities. Finally, the Wasserstein metric is applicable to any location in the flow, thereby providing fine-grained information about the simulation accuracy, model deficiencies in predicting certain scalar quantities, and the impact of inconsistencies of boundary conditions on the simulation. These three fundamental features will be emphasized and built upon as Sec.~\ref{SSEC_MULTI} considers the multidimensional validation of a LES combustion model using experimental data.
\subsection{\label{SSEC_MULTI}Multiscalar results}

Having provided a comparison for the Wasserstein metric applied to mixture fraction as a single scalar, this second test case will demonstrate the application of the Wasserstein metric to multiple scalars in the form of joint scalar distributions. This property allows for multiple, simultaneous error calculations, which provide a multifaceted, quantitative validation. Here, the Wasserstein metric is used to compare simulation results with experimental data for the piloted jet flame with inhomogeneous inlets as discussed in Sec.~\ref{SEC_CONFIG}. 

To quantitatively assess a combustion simulation, a multiscalar Wasserstein metric can be evaluated that takes into account $d$ scalar quantities. In the present cases, four scalar quantities are considered, namely mixture fraction, temperature, and species mass fractions of \ce{CO_2} and \ce{CO}, and evaluations are performed at different axial locations in the jet flame.

\begin{figure}[!htb!]
\begin{center}
 \includegraphics[width = 0.75 \textwidth]{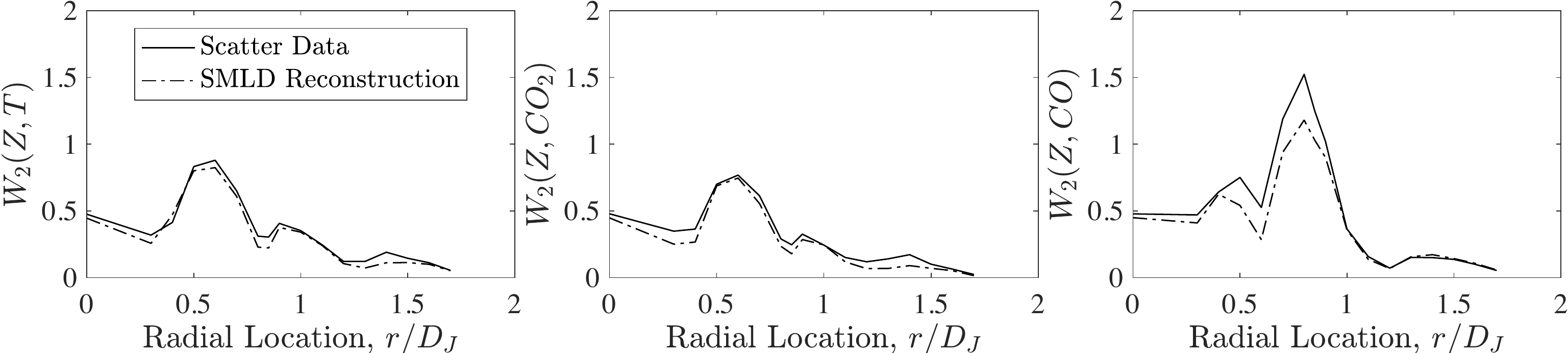}
 \caption{\label{FIG_2D_DEVIATION} Wasserstein metric, evaluated at $x/D_J = 10$: $Z$-$T$, $Z$-$Y_{\ce{CO_2}}$, and $Z$-$Y_{\ce{CO}}$. $W_2$ are calculated based on scatter data (solid line) and SMLD-reconstructed data (dash-dotted line). }
\end{center}
\end{figure}

\begin{figure}[!htb!]
\begin{center}
 \includegraphics[width = \textwidth]{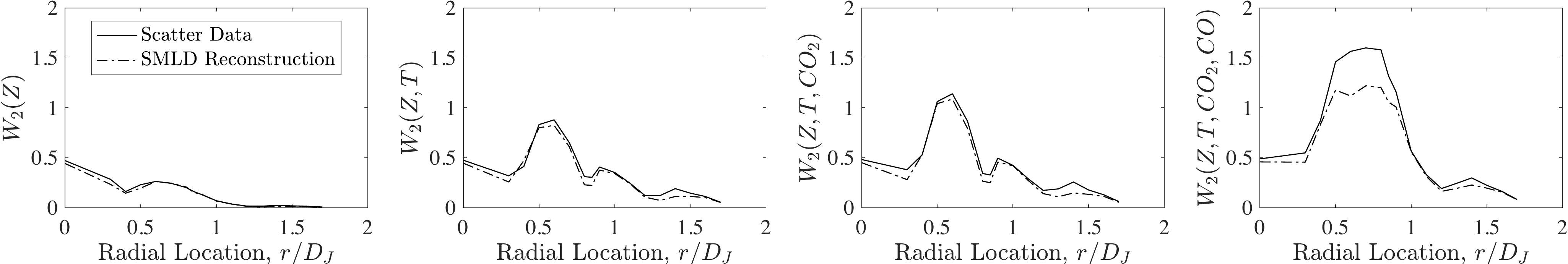}
 \caption{\label{FIG_DEVIATION} Wasserstein metric, evaluated for four scalars at $x/D_J = 10$: $Z$, $Z$-$T$, $Z$-$T$-$Y_{\ce{CO_2}}$, and $Z$-$T$-$Y_{\ce{CO_2}}$-$Y_{\ce{CO}}$. The plots demonstrate deviations in $W_2$ calculations based on data sources: scatter data (solid line) and SMLD-reconstructed data (dash-dotted line). }
\end{center}
\end{figure}

The $W_2$-metric at $x/D_J = 10$ for three two-scalar cases: $Z$-$T$, $Z$-$Y_{\ce{CO_2}}$, and $Z$-$Y_{\ce{CO}}$ are shown in Fig.~\ref{FIG_2D_DEVIATION}. The $W_2$-metric of similar magnitude and radial profile are found in the cases of $Z$-$T$ and $Z$-$Y_{\ce{CO_2}}$, in which there is little discrepancy between the results obtained from the scatter data and SMLD reconstruction. The $Z$-$Y_{\ce{CO}}$ case exhibits a much higher level of $W_2$. There is also greater difference between the values from two sources of experimental results. We then examine how the $W_2$-metric is affected by increasing the number of scalars that is included in its evaluation. For this, we consider results at the same axial location of $x/D_J=10$, and results are presented in Fig.~\ref{FIG_DEVIATION}. The result of $Z$-$Y_{\ce{CO_2}}$ is repeated for clarity. Since the Wasserstein metric measures differences in distribution, it allows for a direct evaluation of how differences arising from uncertainties in boundary conditions or modeling errors manifest in the flow field. The values for the Wasserstein metric increases as more quantities are considered (from left to right in Fig.~\ref{FIG_DEVIATION}). This is to be expected as the pair-wise distances becomes larger with the inclusion of additional dimensions.

A comparison of the Wasserstein metric, evaluated from the SMLD-reconstruction, shows that this statistical reconstruction technique qualitatively and quantitatively captures the results obtained from the scatter data. However, the main deviation made by the SMLD-reconstruction arise when including $Y_{\ce{CO}}$, which indicates deficiencies of the uncorrelated normal distribution in representing the actual joint distribution including $\ce{CO}$ and possible strong correlations between $\ce{CO}$ and other quantities.

\begin{figure}[!htb!]
\begin{center}
 \subfigure[\label{FIG_M_STACK_SCATTER}Scatter data.]{\includegraphics[width = 0.465\textwidth]{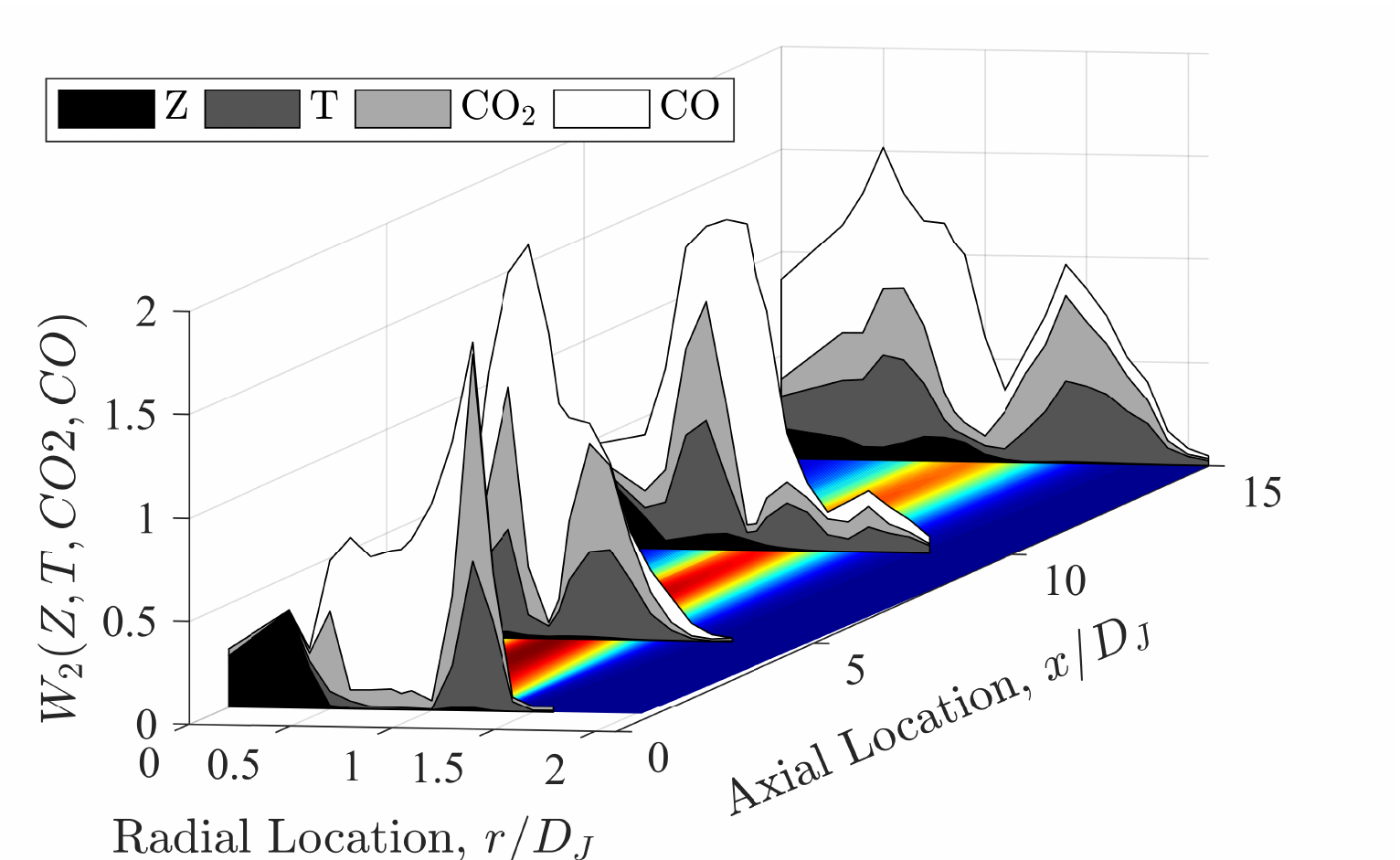}}
 \subfigure[\label{FIG_M_STACK_SMLD}SMLD reconstruction.]{\includegraphics[width = 0.525\textwidth]{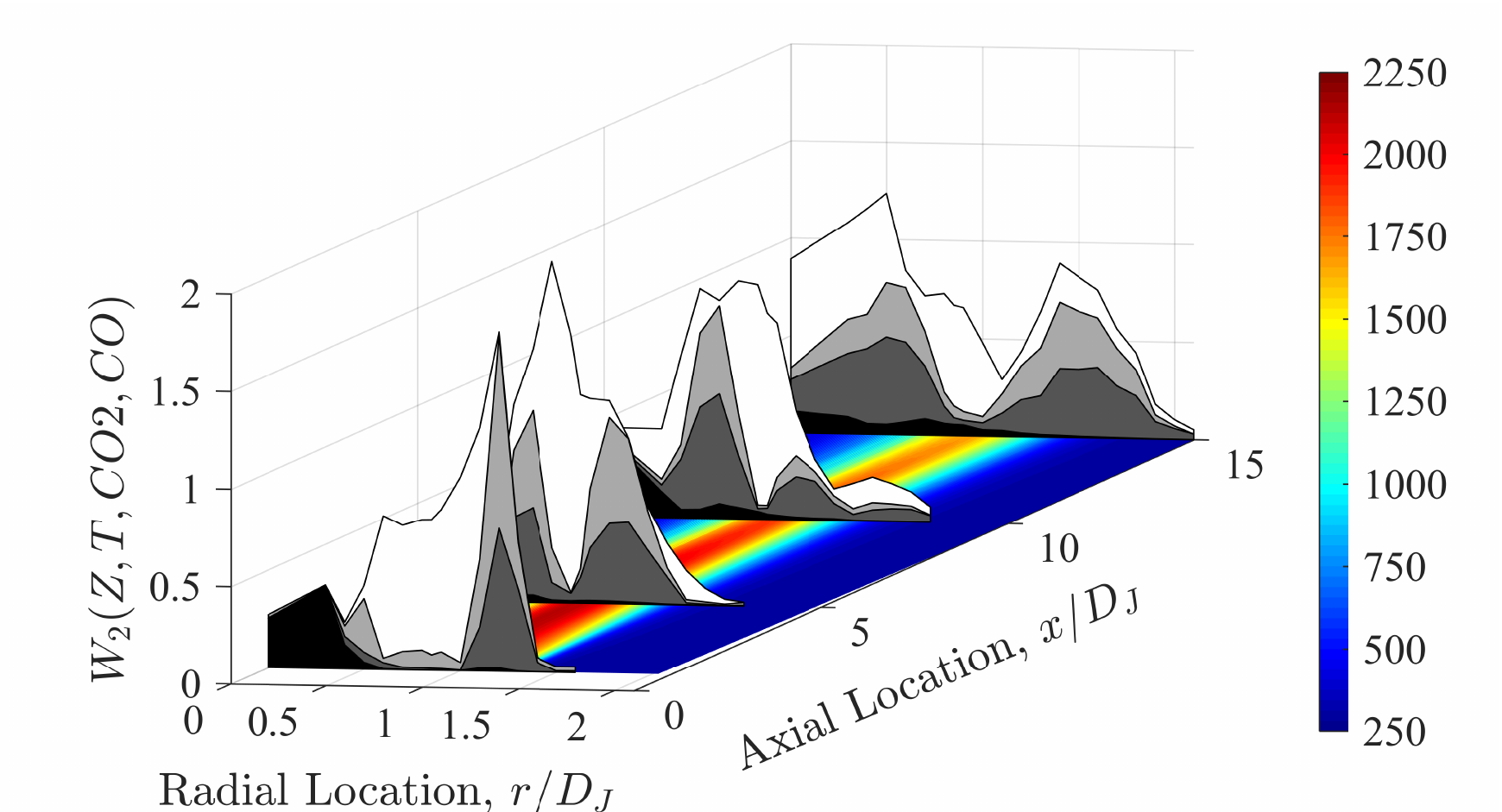}}
 \caption{\label{FIG_M_STACK}Calculations of Wasserstein metric, from (a) scatter data and (b) SMLD-reconstructed data that are evaluated from measurements and LES computations. The stacked results draw attention to the contributions from each dimension of comparison. The $W_2$-metric is calculated along radial profiles at $x/D_J = \{1, 5, 10, 15\}$, and is non-dimensionalized using the standard deviation of experimental data. Superimposed on the floor of each plot are mean experimental temperature values, to illustrate where the error calculations lie within the flame.}
\end{center}
\end{figure}

Having built the multiscalar Wasserstein metric, results for calculations at different axial locations in the flame are displayed in Fig.~\ref{FIG_M_STACK}. The figure on the left represents a $W_2$-comparison made using scatter data, while the figure on the right represents a comparison made using SMLD-reconstructed data. Although scatter data provides a more representative conclusion, SMLD data, reconstructed from mean and variance data, may be more practical based on the existing data reported from simulation results. Since the data has been normalized by the standard deviation, as explained in Sec.~\ref{SSEC_CALC_PROC}, the error metric is unitless. This property, along with the stacking of error contributions, allows for simple comparisons between species, as well as comparisons between axial locations. The stacking represents error contributions that are computed based on each scalar's contribution in pair-wise distances and the transport matrix $\Gamma$ obtained from the four-scalar Wasserstein metric. It can be seen that main contributions to the deviations arise from the mixture fraction at the jet core, and approximately equal distributions from temperature and CO mass fractions in the shear layer. This information can be used to guide corrections in the boundary conditions and the selection of the combustion model. Since the transport matrices $\Gamma$ are different between cases of different dimensionalities, no direct correspondence can be established between the stacking in Fig.~\ref{FIG_M_STACK} and the lower dimensional results in Fig.~\ref{FIG_DEVIATION}. Note also that by increasing the number of scalar quantities in the evaluation of the Wasserstein metric, correlations and scalar interdependencies are taken into account.

Figure~\ref{FIG_M_STACK} reveals the sources and locations of model error. For example, considering $x/D_J = 10$, most of the error is concentrated in the region between $0.5\leq r/D_J \leq 1$. The largest contributors to this error are $Y_{\ce{CO}}$ and $T$. There is additional, but less significant, error in the central jet region from $Z$, which can be attributed to deficiencies in the boundary conditions. Finally, the model error drops off from all sources on the edge of the domain. Similar analyses could be conducted for the other axial locations, but this example demonstrates the usefulness of the Wasserstein metric as a validation tool. Information from these calculations could be used to identify model limitations, isolate regions where mesh refinement is needed, and where further measurements are required.

Three further advantages of using the Wasserstein metric in combustion validation are outlined below. One of the method's benefits is that it isolates sensitivity to boundary conditions. The Wasserstein metric calculations for $x/D_J = 1$ provide some indication of the error at the boundary conditions. As results downstream are examined, it can be seen that the error features from the inlet are convecting and diminishing. One example of this behavior is the mixture fraction error in the jet region, which is a significant source of error for $x/D_J = 1$, but a minor source of error downstream at $x/D_J = 15$. By comparing slices at the inlet and slices downstream, one can identify how boundary conditions introduce errors to the combustion model. Although some error features are diffusing, others are forming downstream. 

A second benefit to this approach is the detection of modeling errors not arising from boundary conditions. These new peaks arise from deficiencies in the combustion model. For instance, the error for $Y_{\ce{CO}}$, just inside $r/D_J =1$ is modest at the inlet, peaks sharply at $x/D_J = 5$, and reduces at $x/D_J = 10$, before peaking sharply again at $x/D_J = 15$. This error represents a deficiency in the CO modeling, and it could be targeted for improvement in future versions of the combustion model. In this way, the Wasserstein metric helps identify regions of modeling error and highlights potential areas for model improvement.

Finally, a third benefit to the Wasserstein metric validation approach is the seamless transition to multiscalar quantities and the contraction of multidimensional information into a single scalar value. Any number of additional species could be added to these plots, without a significant increase in computational cost. The resulting higher-dimensional calculations would offer an even more detailed comparison of the simulated and experimental data, and provide more insight into boundary condition and combustion modeling error. Condensing multiple validation plots into a compact metric makes the validation process easier to interpret, and provides greater understanding about the effectiveness of combustion models.

\section{\label{SEC_TNF}Quantitative comparison of different simulation results}
In this section, we seek to demonstrate the capability of the Wasserstein metric as a tool for quantitative comparisons of simulation results that are generated using different modeling strategies. The direct application of this metric has the potential for eliminating the need for subjective model assessments. It can also provide a direct evaluation of strengths and weaknesses of certain modeling approaches and guides the need for further experimental data to constraint modeling approaches.

In this study, we resort to experimental and computational data of the piloted turbulent burner with inhomogeneous inlets that were collectively reported at the 13th International Workshop on Measurement and Computation of Turbulent Nonpremixed Flames (TNF)~\cite{BARLOWTNF2016}. By complying with the TNF-spirit of open scientific collaborations and acknowledging that results reported at this venue are a work-in-progress, we removed any reference to the original authorship in the following representation. In this context, we would like to emphasize that the present investigation is not intended to provide any judgment about a particular modeling strategy. Such an endeavor requires a concerted community effort, by which the herein proposed Wasserstein metric could come to use as a quantitative measure. It is noted that several of these TNF-contributions have been extended since, and we would like to refer to publications~\cite{WANG_ZIEKER_SCHIESSL_PLATOVA_FROEHLICH_MAAS_PCI2017,PROCH_KEMPF_CF2014,WU_IHME_FUEL2016,WANG_ZHANG_PCI2017,KLEINHEINZ_KUBIS_TRISJONO_BODE_PITSCH_PCI2017,PERRY_MUELLER_MASRI_PCI2017,LUO_YANG_BAI_FAN_FUEL2017}, which provide further details on the validation and analysis of simulation results, as well as description of combustion models and computational setups of individual groups.

%
In this study, we concentrate on the flame configuration {FJ-5GP-Lr75-57} that was discussed in Sec.~\ref{SSEC_CONFIG_EXP} and was selected as a target configuration at the TNF-workshop. The following analysis concentrates on comparisons of scatter plots of mixture fraction, temperature, and species mass fractions of \ce{CO2} and \ce{CO} at four axial locations $x/D_J=\{1,5,10,15\}$. 

\begin{figure}[!t!]
\begin{center}
 \includegraphics[width = \textwidth]{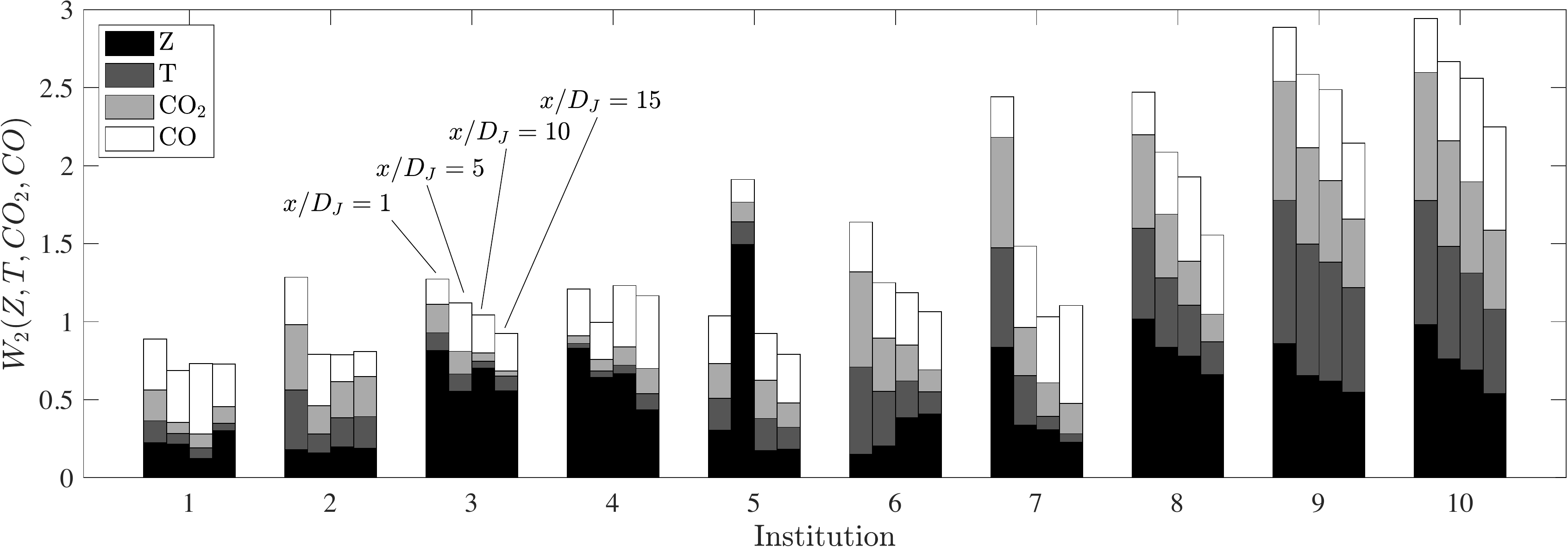}
 \caption{\label{FIG_UNIV_COMPARE}Quantitative comparison of multiscalar Wasserstein metric, $W_2(Z,T,Y_{\ce{CO2}}, Y_{\ce{CO}})$ from ten anonymized LES-calculations, presented at the 13th TNF-workshop~\cite{BARLOWTNF2016}. The decomposition of multiscalar calculations allows contributions from each variable in each axial location to become visible. This quantitative validation analysis enables models to be compared objectively. The four bar-graphs from each contribution correspond to axial locations of $x/D_J=\{1,5,10,15\}$. (Results used in this figure were included with permission from TNF-contributors.)}
\end{center}
\end{figure}

In order to apply the Wasserstein metric to this extensive set of modeling results, we proceeded as follows. First, we downsampled the scatter data by randomly selecting 5000 points, each point containing information for $Z,T,Y_{\ce{CO2}},$ and $Y_{\ce{CO}}$. This downsampling procedure was performed to achieve reasonable runtimes. The chosen number of samples was found to be appropriate for representing the simulation results without loss of information, and a $5\%$ level of uncertainty is estimated for the $W_p$ of multiscalar cases. Next, each data set was normalized by the experimental standard deviation, as set forth in Sec.~\ref{SSEC_CALC_PROC}. Subsequently, the multiscalar Wasserstein metric was evaluated from each data set, and results are presented in Fig.~\ref{FIG_UNIV_COMPARE}. 

From this direct comparison, the following observations can be drawn. First, the multiscalar Wasserstein metric, $W_2(Z,T,Y_{\ce{CO2}}, Y_{\ce{CO}})$, shows cumulative contributions of each scalar quantity to the overall error that is invoked by the model prediction. As discussed in Sec.~\ref{SSEC_WASSERSTEIN_METRIC}, the analysis is performed using properly non-dimensionalized quantities, so that the $W_2$-metric provides a direct representation of the error; a numerical value of $W_2=1$ corresponds to an error of one standard deviation with respect to the measurement. From this follows the second observation that this decomposition of the $W_2$-metric provides direct information about relative contributions of each quantity that is included in its construction. Since the Wasserstein metric is constructed from joint scatter data, it takes into account correlations among scalar quantities, and can therefore be employed in isolating causalities in model predictions. For instance, it can be seen that simulations from Institutions 1 and 2 exhibit a relatively small error contribution from mixture fraction and temperature, but a larger contribution from CO mass fraction. In contrast, Institutions 3 and 4 show a bias towards larger errors in the mixture fraction data, whereas Institutions 9 and 10 show an approximately equally distributed contribution from all four quantities to the $W_2$-metric. Such quantitative information can be useful in isolating model deficiencies and guiding the formulation of model extensions. 

A third observation can be drawn by considering the axial evolution of the $W_2$-metric. Specifically, from Fig.~\ref{FIG_UNIV_COMPARE} it can be seen that most simulations show a reduction in the error with increasing axial distance. This trend can be explained by the equilibration of the combustion, reduction of the mixing, and decay of the turbulence, which is typical for canonical jet flames. In turn, significant deviations from this anticipated trend can hint at potential deficiencies in the model setup, or physically interesting combustion behavior (such as local extinction, multistream mixing, flame lifting, or local vortex break-down) that requires further experimental investigation or refinement of the modeling procedure, improving mesh-resolution, or adjustments of boundary conditions. As example, the simulation from Institution 5 in Fig.~\ref{FIG_UNIV_COMPARE} shows significant deviation of the mixture fraction at the second measurement location ($x/D_J=5$), suggesting that results at this measurement location require further analysis. 

The fourth observation arises from the utility of the Wasserstein metric in separating contributions from boundary conditions and combustion-model formulations. For instance, dominant discrepancies at the first measurement location, $x/D_J=1$, can most likely be attributed to uncertainties in the boundary conditions. Contribution of different combustion models can then be tested to examine model developments, and guide model selections for simulating particular flame configurations. Simulation results reported from this comparative study show a rather wide variation in model accuracy, ranging between $W_2=0.8$ for the most accurate simulation to values of $W_2=3$, with largest errors occurring at the first reported measurement station. 

These results show that the application of the Wasserstein metric to a large set of simulation data provides for a quantitative assessment of different simulation results. The practical utility of this metric lies in the direct assessment of the model performance and in tracking the model convergence as continuous improvements on the model formulation, boundary conditions, and measurements are provided. In regard to providing a verbal evaluation of the accuracy of a particular model, the Wasserstein metric can be utilized as practically useful model performance index for model ranking~\cite{WILLMOTT_BAMS1982}. The Wasserstein metric provides a way of ranking models, which we believe is an initial step towards advancing progress in combustion modeling.
\section{\label{SEC_CONCLUSION}Conclusions}
This manuscript addresses the need for the validation of numerical simulations by considering multiple scalar quantities (velocity, temperature, composition), different data presentations (statistical results, scatter data, and conditional results), and various data acquisitions (pointwise, 1D, and planar measurements at different spatial locations). To this end, the Wasserstein metric is proposed as a general formulation to facilitate quantitative and objective comparisons between measurements and simulations. This metric is a probabilistic measure and is applicable to empirical distributions that are generated from scatter data or statistical results using probabilistic reconstruction. 

The Wasserstein metric was derived for turbulent-combustion applications and essential convergence properties where examined. The resulting method was demonstrated in application to LES of a partially premixed turbulent jet flame, and used to categorize errors arising from deficiencies in the specification of boundary conditions and intrinsic limitations of combustion models. This investigation was followed by applying the Wasserstein metric to different simulations that were contributed to the TNF-workshop in order to demonstrate the versatility of this method in establishing an objective evaluation of large data sets.

In building upon previous statistical validation approaches, the Wasserstein metric offers greater insight by condensing multiscalar differences between data into a single error measure. The Wasserstein metric gives an accurate indication of errors over the whole flame, and its interpretation can assist in identifying causalities of discrepancies in numerical simulation, arising from boundary conditions, complex flow-field structures, or model limitations. The method is easy to apply, and its major benefit lies in the seamless extension to multiscalar analyses, thereby taking into account interdependencies of combustion-physical processes. When applied to combustion validation, the Wasserstein metric can concisely communicate shortcomings of current models, and assist in the development of more accurate LES combustion models.
\section*{Acknowledgments}
Financial support through NASA's Transformational Tools and Technologies Program with Award No. NNX15AV04A is gratefully acknowledged. Resources supporting this work were provided by the NASA High-End Computing (HEC) Program through the NASA Advanced Supercomputing (NAS) Division at Ames Research Center and the National Energy Research Scientific Computing Center, a DOE Office of Science User Facility supported by the Office of Science of the U.S. Department of Energy under Contract No. DE-AC02-05CH11231. We would like to thank Benoit Fiorina, M\'{e}lody Cailler, and contributors to the 2016 TNF Workshop for permitting us to use the computational data presented in Fig.~\ref{FIG_UNIV_COMPARE} for analysis. To acknowledge that these results were a work-in-progress and honor TNF-policy, we anonymized individual contributions.
\bibliographystyle{unsrt}
\bibliography{combustion,emd,stats}

\begin{thebibliography}{10}

\bibitem{HEITOR_MOREIRA_PECS1993}
M.~V. Heitor and A.~L.~N. Moreira.
\newblock Thermoacouples and sample probes for combustion studies.
\newblock {\em Prog. Energy Combust. Sci.}, 19:259--278, 1993.

\bibitem{ECKBRETH_BOOK1996}
A.~C. Eckbreth.
\newblock {\em Laser Diagnostics for Combustion Temperature and Species}.
\newblock Gordon and Breach, 1996.

\bibitem{KOHSE_HOINGHAUS_BARLOW_ALDEN_WOLFRUM_PCI2005}
K.~Kohse-H\"oinghaus, R.~S. Barlow, M.~Ald\'en, and J.~Wolfrum.
\newblock Combustion at the focus: laser diagnostics and control.
\newblock {\em Proc. Combust. Inst.}, 30(1):89--123, 2005.

\bibitem{BARLOW_PCI2007}
R.~S. Barlow.
\newblock Laser diagnostics and their interplay with computations to understand
  turbulent combustion.
\newblock {\em Proc. Combust. Inst.}, 31:49--75, 2007.

\bibitem{BARLOW_FRANK_KARPETIS_CHEN_CF2005}
R.~S. Barlow, J.~H. Frank, A.~N. Karpetis, and J.-Y. Chen.
\newblock Piloted methane/air jet flames: {Transport} effects and aspects of
  scalar structure.
\newblock {\em Combust. Flame}, 143:433--449, 2005.

\bibitem{ALDEN_BOOD_LO_RICHTER_PCI2011}
M.~Ald\'en, J.~Bood, Z.~Li, and M.~Richter.
\newblock Visualization and understanding of combustion processes using
  spatially and temporally resolved laser diagnostic techniques.
\newblock {\em Proc. Combust. Inst.}, 33(1):69--97, 2011.

\bibitem{BARLOWTNF}
R.~S. Barlow, 1996.
\newblock Web site for the International Workshop on Measurement and
  Computation of Turbulent Nonpremixed Flames (TNF),
  \url{http://www.ca.sandia.gov/TNF/abstract.html}.

\bibitem{KEMPF_GEURTS_OEFELEIN_CF2011}
A.~M. Kempf, B.~J. Geurts, and J.~C. Oefelein.
\newblock Error analysis of large-eddy simulation of the turbulent non-premixed
  {Sydney} bluff-body flame.
\newblock {\em Combust. Flame}, 158(12):2408--2419, 2011.

\bibitem{KHALIL_LACAZE_OEFELEIN_NAJM_PCI2015}
M.~Khalil, G.~Lacaze, J.~C. Oefelein, and H.~N. Najm.
\newblock Uncertainty quantification in {LES} of a turbulent bluff-body
  stabilized flame.
\newblock {\em Proc. Combust. Inst.}, 35(2):1147--1156, 2015.

\bibitem{BOGACHEV_BOOK2007}
V.~I. Bogachev.
\newblock {\em Measure Theory}.
\newblock Springer, 2007.

\bibitem{gibbs2002choosing}
A.~L. Gibbs and F.~E. Su.
\newblock On choosing and bounding probability metrics.
\newblock {\em International statistical review}, 70(3):419--435, 2002.

\bibitem{Rubner2000}
Y.~Rubner, C.~Tomasi, and L.~J. Guibas.
\newblock Earth mover's distance as a metric for image retrieval.
\newblock {\em Int. J. Comput. Vision}, 40(2):99--121, 2000.

\bibitem{ren2011robust}
Z.~Ren, J.~Yuan, and Z.~Zhang.
\newblock Robust hand gesture recognition based on finger-earth mover's
  distance with a commodity depth camera.
\newblock In {\em Proceedings of the 19th ACM international conference on
  Multimedia}, pages 1093--1096. ACM, 2011.

\bibitem{su2015optimal}
Z.~Su, Y.~Wang, R.~Shi, W.~Zeng, J.~Sun, F.~Luo, and X.~Gu.
\newblock Optimal mass transport for shape matching and comparison.
\newblock {\em IEEE transactions on pattern analysis and machine intelligence},
  37(11):2246--2259, 2015.

\bibitem{de2011optimal}
F.~de~Goes, D.~Cohen-Steiner, P.~Alliez, and M.~Desbrun.
\newblock An optimal transport approach to robust reconstruction and
  simplification of 2d shapes.
\newblock {\em Computer Graphics Forum}, 30(5):1593--1602, 2011.

\bibitem{villani2008optimal}
C.~Villani.
\newblock {\em Optimal Transport: Old and New}, volume 338 of {\em A Series of
  Comprehensive Studies in Mathematics}.
\newblock Springer, 2008.

\bibitem{urbas1998mass}
J.~Urbas.
\newblock {\em Mass Transfer Problems}, volume~41 of {\em
  Sonderforschungsbereich Nichtlineare Partielle Differentialgleichungen Bonn:
  Vorlesungsreihe}.
\newblock Sonderforschungsbereich 256, 1998.

\bibitem{mccann2011five}
R.~J. McCann and N.~Guillen.
\newblock Five lectures on optimal transportation: geometry, regularity and
  applications.
\newblock {\em Analysis and geometry of metric measure spaces: lecture notes of
  the s{\'e}minaire de Math{\'e}matiques Sup{\'e}rieure (SMS) Montr{\'e}al},
  pages 145--180, 2011.

\bibitem{monge1781memoire}
G.~Monge.
\newblock {\em M{\'e}moire sur la th{\'e}orie des d{\'e}blais et des remblais}.
\newblock De l'Imprimerie Royale, 1781.

\bibitem{kuhn1955hungarian}
H.~W. Kuhn.
\newblock The hungarian method for the assignment problem.
\newblock {\em Naval research logistics quarterly}, 2(1-2):83--97, 1955.

\bibitem{kantorovich1958space}
L.~V. Kantorovich and G.~S. Rubinstein.
\newblock On a space of completely additive functions.
\newblock {\em Vestnik Leningrad. Univ}, 13(7):52--59, 1958.

\bibitem{kantorovich2006problem}
L.~V. Kantorovich.
\newblock On a problem of monge.
\newblock {\em J. Math. Sci.}, 133(4):1383--1383, 2006.

\bibitem{dudley1976probabilities}
R.~M. Dudley.
\newblock {\em Probabilities and metrics: Convergence of laws on metric spaces,
  with a view to statistical testing}, volume~45.
\newblock Aarhus Universitet, Matematisk Institut, 1976.

\bibitem{POPE_BOOK00}
S.~B. Pope.
\newblock {\em Turbulent Flows}.
\newblock Cambridge University Press, Cambridge, 2000.

\bibitem{RUBINSTEIN_KROESE_BOOK2008}
R.~Y. Rubinstein and D.~P. Kroese.
\newblock {\em Simulation and the Monte Carlo method}.
\newblock John Wiley \& Sons, Inc., 2008.

\bibitem{bickel1981some}
P.~J. Bickel and D.~A. Freedman.
\newblock Some asymptotic theory for the bootstrap.
\newblock {\em Ann. Stat.}, pages 1196--1217, 1981.

\bibitem{horowitz1994mean}
J.~Horowitz and R.~L. Karandikar.
\newblock Mean rates of convergence of empirical measures in the {Wasserstein}
  metric.
\newblock {\em Jour. Comp. App. Math.}, 55(3):261--273, 1994.

\bibitem{fournier2015rate}
N.~Fournier and A.~Guillin.
\newblock On the rate of convergence in wasserstein distance of the empirical
  measure.
\newblock {\em Probab. Theory Relat. Fields}, 162(3-4):707--738, 2015.

\bibitem{efron1979bootstrap}
B.~Efron.
\newblock Bootstrap methods: another look at the jackknife.
\newblock {\em Ann. Stat.}, pages 1--26, 1979.

\bibitem{chernick2011bootstrap}
M.~R. Chernick, W.~Gonz{\'a}lez-Manteiga, R.~M. Crujeiras, and E.~B. Barrios.
\newblock {\em Bootstrap methods}.
\newblock Springer, 2011.

\bibitem{bickel1996m}
P.~J. Bickel and J.~J. Ren.
\newblock The m out of n bootstrap and goodness of fit tests with double
  censored data.
\newblock In {\em Robust Statistics, Data Analysis, and Computer Intensive
  Methods}, pages 35--47. Springer, 1996.

\bibitem{sommerfeld2016inference}
M.~Sommerfeld and A.~Munk.
\newblock Inference for empirical wasserstein distances on finite spaces.
\newblock {\em arXiv preprint arXiv:1610.03287}, 2016.

\bibitem{dumbgen1993nondifferentiable}
L.~D{\"u}mbgen.
\newblock On nondifferentiable functions and the bootstrap.
\newblock {\em Probab. Theory Related Fields}, 95(1):125--140, 1993.

\bibitem{POPE_JNT79}
S.~B. Pope.
\newblock A rational method of determining probability distributions in
  turbulent reacting flows.
\newblock {\em J. Non-Equilib. Thermodyn.}, 4:309--320, 1979.

\bibitem{IHME_PITSCH_PRT1_CF_2008}
M.~Ihme and H.~Pitsch.
\newblock Prediction of extinction and reignition in non-premixed turbulent
  flames using a flamelet/progress variable model. {1.} {A} priori study and
  presumed {PDF} closure.
\newblock {\em Combust. Flame}, 155:70--89, 2008.

\bibitem{Coclite2015}
A.~Coclite, G.~Pascazio, P.~De~Palma, L.~Cutrone, and M.~Ihme.
\newblock An {SMLD} joint pdf model for turbulent non-premixed combustion using
  the flamelet progress-variable approach.
\newblock {\em Flow, Turb. Combust.}, 95(1):97--119, 2015.

\bibitem{lisman1972note}
J.~H.~C. Lisman and M.~C.~A. Van~Zuylen.
\newblock Note on the generation of most probable frequency distributions.
\newblock {\em Stat. Neerl.}, 26(1):19--23, 1972.

\bibitem{Pele2009}
O.~Pele and M.~Werman.
\newblock {Fast and robust earth mover's distances}.
\newblock {\em Proceedings of the IEEE International Conference on Computer
  Vision}, pages 460--467, 2009.

\bibitem{ahuja1988network}
R.~K. Ahuja.
\newblock {\em Network Flows}.
\newblock PhD thesis, Technische Hochschule Darmstadt, 1988.

\bibitem{muskulus2011wasserstein}
M.~Muskulus and S.~Verduyn-Lunel.
\newblock Wasserstein distances in the analysis of time series and dynamical
  systems.
\newblock {\em Physica D}, 240(1):45--58, 2011.

\bibitem{doi:10.1093/biomet/asv068}
B.~Efron.
\newblock Empirical bayes deconvolution estimates.
\newblock {\em Biometrika}, 103(1):1, 2016.

\bibitem{laird1978nonparametric}
N.~Laird.
\newblock Nonparametric maximum likelihood estimation of a mixing distribution.
\newblock {\em JASA}, 73(364):805--811, 1978.

\bibitem{MEARES_MASRI_CF2014}
S.~Meares and A.~R. Masri.
\newblock A modified piloted burner for stabilizing turbulent flames of
  inhomogeneous mixtures.
\newblock {\em Combust. Flame}, 161(2):484--495, 2014.

\bibitem{MEARES_PRASAD_MAGNOTTI_BARLOW_MASRI_PCI2015}
S.~Meares, V.~N. Prasad, G.~Magnotti, R.~S. Barlow, and A.~R. Masri.
\newblock Stabilization of piloted turbulent flames with inhomogeneous inlets.
\newblock {\em Proc. Combust. Inst.}, 35(2):1477--1484, 2015.

\bibitem{BARLOW_MEARES_MAGNOTTI_CUTCHER_MASRI_CF2015}
R.~S. Barlow, S.~Meares, G.~Magnotti, H.~Cutcher, and A.~R. Masri.
\newblock Local extinction and near-field structure in piloted turbulent
  {CH$_4$/air} jet flames with inhomogeneous inlets.
\newblock {\em Combust. Flame}, 162(10):3516--3540, 2015.

\bibitem{moin2006large}
P.~Moin and S.~V. Apte.
\newblock Large-eddy simulation of realistic gas turbine combustors.
\newblock {\em AIAA J.}, 44(4):698--708, 2006.

\bibitem{esclapez2017fuel}
L.~Esclapez, P.~C. Ma, E.~Mayhew, R.~Xu, S.~Stouffer, T.~Lee, H.~Wang, and
  M.~Ihme.
\newblock Fuel effects on lean blow-out in a realistic gas turbine combustor.
\newblock {\em Combust. Flame}, 181:82--99, 2017.

\bibitem{CHUCK_JFM04}
C.~D. Pierce and P.~Moin.
\newblock Progress-variable approach for large-eddy simulation of non-premixed
  turbulent combustion.
\newblock {\em J. Fluid Mech.}, 504:73--97, 2004.

\bibitem{IHME_PCI04}
M.~Ihme, C.~M. Cha, and H.~Pitsch.
\newblock Prediction of local extinction and re-ignition effects in
  non-premixed turbulent combustion using a flamelet/progress variable
  approach.
\newblock {\em Proc. Combust. Inst.}, 30:793--800, 2005.

\bibitem{PETERS_PECS84}
N.~Peters.
\newblock Laminar diffusion flamelet models in non-premixed turbulent
  combustion.
\newblock {\em Prog. Energy Combust. Sci.}, 10(3):319--339, 1984.

\bibitem{POINSOTBOOK01}
T.~Poinsot and D.~Veynante.
\newblock {\em Theoretical and Numerical Combustion}.
\newblock R.T. Edwards, Inc., Philadelphia, PA, 2001.

\bibitem{IHME_PITSCH_PRT2_CF_2008}
M.~Ihme and H.~Pitsch.
\newblock Prediction of extinction and reignition in non-premixed turbulent
  flames using a flamelet/progress variable model. {2.} {A} posteriori study
  with application to {Sandia} flames {D} and {E}.
\newblock {\em Combust. Flame}, 155:90--107, 2008.

\bibitem{WU_IHME_FUEL2016}
H.~Wu and M.~Ihme.
\newblock Compliance of combustion models for turbulent reacting flow
  simulations.
\newblock {\em Fuel}, 186:853¿863, 2016.

\bibitem{CHEN_IHME_CF2013}
Y.~Chen and M.~Ihme.
\newblock Large-eddy simulation of a piloted premixed jet burner.
\newblock {\em Combust. Flame}, 160:2896--2910, 2013.

\bibitem{IHME_SEE_PCI33}
M.~Ihme and Y.~C. See.
\newblock {LES} flamelet modeling of a three-stream {MILD} combustor: Analysis
  of flame sensitivity to scalar inflow conditions.
\newblock {\em Proc. Combust. Inst.}, 33, 2010.
\newblock 1309-1317.

\bibitem{WU_SEE_WANG_IHME_CF2015}
H.~Wu, Y.~C. See, Q.~Wang, and M.~Ihme.
\newblock A {Pareto-efficient} combustion framework with submodel assignment
  for predicting complex flame configurations.
\newblock {\em Combust. Flame}, 162:4208--4230, 2015.

\bibitem{wang2017regularized}
Q.~Wang and M.~Ihme.
\newblock Regularized deconvolution method for turbulent combustion modeling.
\newblock {\em Combust. Flame}, 176:125--142, 2017.

\bibitem{Cook1994}
A.~W. Cook and J.~J. Riley.
\newblock {A subgrid model for equilibrium chemistry in turbulent flows}.
\newblock {\em Physics of Fluids}, 6(8):2868, 1994.

\bibitem{rice2006mathematical}
J.~Rice.
\newblock {\em Mathematical Statistics and Data Analysis}.
\newblock Nelson Education, 2006.

\bibitem{BARLOWTNF2016}
R.~S. Barlow, 2016.
\newblock 13th International Workshop on Measurement and Computation of
  Turbulent Nonpremixed Flames (TNF),
  \url{http://www.sandia.gov/TNF/13thWorkshop/TNF13.html}.

\bibitem{WANG_ZIEKER_SCHIESSL_PLATOVA_FROEHLICH_MAAS_PCI2017}
P.~Wang, F.~Zieker, R.~Schie{\ss}l, N.~Platova, J.~Fr\"ohlich, and U.~Maas.
\newblock Large eddy simulations and experimental studies of turbulent premixed
  combustion near extinction.
\newblock {\em Proc. Combust. Inst.}, 34(1):1269--1280, 2013.

\bibitem{PROCH_KEMPF_CF2014}
F.~Proch and A.~M. Kempf.
\newblock Numerical analysis of the cambridge stratified flame series using
  artificial thickened flame {LES} with tabulated premixed flame chemistry.
\newblock {\em Combust. Flame}, 161(10):2627--2646, 2014.

\bibitem{WANG_ZHANG_PCI2017}
H.~Wang and P.~Zhang.
\newblock A unified view of pilot stabilized turbulent jet flames for model
  assessment across different combustion regimes.
\newblock {\em Proc. Combust. Inst.}, 36(2):1693--1703, 2017.

\bibitem{KLEINHEINZ_KUBIS_TRISJONO_BODE_PITSCH_PCI2017}
K.~Kleinheinz, T.~Kubis, P.~Trisjono, M.~Bode, and H.~Pitsch.
\newblock Computational study of flame characteristics of a turbulent piloted
  jet burner with inhomogeneous inlets.
\newblock {\em Proc. Combust. Inst.}, 36(2):1747--1757, 2017.

\bibitem{PERRY_MUELLER_MASRI_PCI2017}
B.~A. Perry, M.~E. Mueller, and A.~R. Masri.
\newblock A two mixture fraction flamelet model for large eddy simulation of
  turbulent flames with inhomogeneous inlets.
\newblock {\em Proc. Combust. Inst.}, 36(2):1767--1775, 2017.

\bibitem{LUO_YANG_BAI_FAN_FUEL2017}
K.~Luo, J.~Yang, Y.~Bai, and J.~Fan.
\newblock Large eddy simulation of turbulent combustion by a dynamic
  second-order moment closure model.
\newblock {\em Fuel}, 187:457--467, 2017.

\bibitem{WILLMOTT_BAMS1982}
C.~J. Willmott.
\newblock Some comments on the evaluation of model performance.
\newblock {\em Bull. Am. Meteorol. Soc.}, 63(11):1309--1313, 1982.

\bibitem{bobkov2014one}
S.~Bobkov and M.~Ledoux.
\newblock One-dimensional empirical measures, order statistics and
  {Kantorovich} transport distances.
\newblock {\em preprint}, 2016.

\bibitem{Cohen1997}
S.~D. Cohen and L.~J. Guibas.
\newblock The earth mover's distance: {Lower} bounds and invariance under
  translation.
\newblock {CS-TR-97-1597}, Stanford University, Department of Computer Science,
  1997.

\end{thebibliography}
\appendix
\section{\label{APP_WM_GEN_PROB_MEASURE}Wasserstein metric for general probability measures}
Let $(M,\, d)$ denote a complete separable metric space equipped with distance functions $d$, on which there are two probability measures $\mu$ and $\nu$ with finite $p^{\text{th}}$ moments. Following Kantorovich's formulation, we have the $p^{\text{th}}$ Wasserstein metric between $\mu$ and $\nu$ defined as
\begin{equation}
\label{eq_general_wasserstein_metric}
	W_p(\mu,\,\nu) = \bigg( \inf_{\gamma \in \Gamma(\mu,\,\nu) } \int_{M \times M} d(x,\,y)^p \, d\gamma(x,\,y) \bigg)^{1/p},\,
\end{equation}
where $ \Gamma(\mu,\,\nu) $ is a set of admissible measures $\gamma$ on $M \times M$, whose marginals are $\mu$ and $\nu$.

Specifically for continuous distributions defined on the Euclidean space with $M = \mathbf{R}^d$ and $d(x,\,y) = |x-y|$, the measures $\mu$ and $\nu$ can be represented by their probability density functions denoted by $f$ and $g$. Thus, we can rewrite the definition in Eq.~\ref{eq_general_wasserstein_metric} as
\begin{equation}
\label{eq_euclidiean_wasserstein_metric}
	W_p(\mu,\,\nu) = \bigg( \inf_{h \in G(f,\,g)} \int_{\mathbf{R}^d} \int_{\mathbf{R}^d} d(x,\,y)^p h(x,\,y) \, dx \, dy \bigg)^{1/p},\,
\end{equation}
where $G(f,\,g)$ is a set of joint probability density functions, whose marginal density functions satisfy
\begin{align}
	\int_{\mathbf{R}^d} h(x,\,y) \, dy &= f(x) \, ,\\
	\int_{\mathbf{R}^d} h(x,\,y) \, dx &= g(y)\,.
\end{align}

The formulation in Eq.~\ref{eq_euclidiean_wasserstein_metric} of the Wasserstein metric entails two different interpretations. The first interpretation is very close to the origin of the optimal transportation problem. If each distribution is viewed as a pile of ``dirt" distributed in the Euclidean space according to the probability, the metric is the minimum amount of work required to turn one pile into the other. The transport plan $h(x,\, y)$ represents the density of mass to be transported from $x$ to $y$. The second interpretation views $h(x,\, y)$ as the joint distribution of $x$ and $y$ whose marginal distributions match $f$ and $g$. The Wasserstein metric is the minimal expectation of the distance between $x$ and $y$ among all such distributions. 

For the special case of one-dimensional distributions on the real line, the Wasserstein metric possesses many useful properties~\cite{bobkov2014one, Cohen1997}. Let $F$ and $G$ be the cumulative distribution functions for one-dimensional distributions $\mu$ and $\nu$, while $F^{-1}$ and $G^{-1}$ being their corresponding inversions. The Wasserstein metric can then be written in explicit form as
\begin{equation}
\label{eq:w_p_cdf}
	W_p(\mu,\, \nu) = \bigg(\int_{0}^{1} | F^{-1}(x) - G^{-1}(x) |^p dx \bigg)^{1/p}\,.
\end{equation}
Furthermore, when $\mu$ and $\nu$ are marginal empirical distributions with the same number of samples, the relationship in Eq.~\ref{eq:w_p_cdf} can be further simplified as 
\begin{equation}
\label{eq:w_p_edf}
	W_p(\mu,\, \nu) = \bigg( \frac{1}{n} \sum_i^n |x_i^* - y_i^*|^p \bigg)^{1/p}\,,
\end{equation}
where $x^*_i$ and $y^*_i$ are $x_i$ and $y_i$ in sorted order.

\section{\label{APP_SRC_CODE}Sample code for the evaluation of the multidimensional Wasserstein metric}
Sample code is provided for three different test-cases, involving the evaluation of the Wasserstein metric for a single scalar quantity ($Z$), joint scalars ($Z$-$T$), and multi scalars ($Z$-$T$-$Y_{\ce{CO2}}$-$Y_{\ce{CO}}$). The code, available at: {\url{https://github.com/IhmeGroup/WassersteinMetricSample}}, is easily adaptable to other conditions. The implementation of the sample code mirrors the procedure laid out in Section~\ref{SSEC_CALC_PROC}. The script \texttt{sampleWasserstein.m} is the main program, while \texttt{calcW2.m} is the function that calculates the Wasserstein metric.

Inputs to \texttt{calcW2.m} include experimental and simulated data samples for the Sydney piloted jet flame, at $x/D_J = 10$, $r/D_J =0.6$, as well as a list of comparison species. Outputs include the calculated Wasserstein metric ($W_2$), transport matrix, and data visualizations (for one-dimensional and two-dimensional cases, only). Additional cases can be calculated by modifying the \texttt{species\_index} input, where the available species include $Z$, $T$, $Y_{\ce{CO_{2}}}$, and $Y_{\ce{CO}}$.

\end{document}